\PassOptionsToPackage{dvipsnames}{xcolor}
\documentclass[twocolumn]{aastex631}

\hypersetup{linkcolor=red,citecolor=teal,filecolor=cyan,urlcolor=magenta}

\def\kms{\ifmmode \mbox{\rm km\ s}^{-1}
	\else {\rm km\ s}$^{-1}$\ignorespaces         
	\fi}
\def\mum{\ifmmode \mu\mbox{\rm m}  
	\else $\mu${\rm m}         
	\fi}
\def\Htwo{\ifmmode \mbox{\rm H}_2   
	\else {\rm H}$_2$         
	\fi}
\def\OHp{\ifmmode \mbox{\rm OH}^+   
	\else {\rm OH$^+$}         
	\fi}
\def\H2Op{\ifmmode \mbox{\rm H}_2{\rm O}^+   
	\else {\rm H}_2{\rm O}$^+$         
	\fi}
\def\CHp{\ifmmode \mbox{\rm CH}^+   
	\else {\rm CH}$^+$         
	\fi}
\def\Hp{\ifmmode \mbox{\rm H}^+   
	\else {\rm H}$^+$         
	\fi}
\def\Hthreep{\ifmmode \mbox{\rm H}^+_3   
	\else {\rm H}$^+_3$         
	\fi}
\def\mHI{\relax                                      
	\ifmmode {\rm m_{\mbox{\scriptsize\rm H\,\sc I}}} 
	\else $m_{\mbox{\scriptsize\rm H\,\sc I}}$
	\fi}
\DeclareRobustCommand{\HI}{%
	\mbox{H\check@mathfonts\fontsize\sf@size\z@\selectfont I}%
}

\accepted{December 18, 2022}

\shorttitle{Molecular Outflows in $z>6$ QSO Hosts}
\shortauthors{Butler et al. 2022}

\watermark{}

\graphicspath{{./}{ }}

\begin{document}

\title{Molecular Outflows in $z>6$ Unobscured QSO Hosts Driven by Star Formation}

\correspondingauthor{Kirsty May Butler}
\email{kirstymaybutler@gmail.com}

\author[0000-0001-7387-0558]{Kirsty M. Butler}
\affiliation{Institut de Radioastronomie Millimétrique (IRAM), 300 rue de la Piscine, 38400 Saint-Martin-d’Hères, France}
\affiliation{Leiden Observatory, Leiden University, PO Box 9513, 2300 RA Leiden, the Netherlands}

\author{Paul P. van der Werf}
\affiliation{Leiden Observatory, Leiden University, PO Box 9513, 2300 RA Leiden, the Netherlands}
	
\author{Theodoros Topkaras}
\affiliation{Leiden Observatory, Leiden University, PO Box 9513, 2300 RA Leiden, the Netherlands}
	
\author{Matus Rybak}
\affiliation{THz Sensing Group, Faculty of Electrical Engineering, Mathematics and Computer Science, TU Delft, The Netherlands}
\affiliation{Leiden Observatory, Leiden University, PO Box 9513, 2300 RA Leiden, the Netherlands}	

\author{Bram P. Venemans}
\affiliation{Leiden Observatory, Leiden University, PO Box 9513, 2300 RA Leiden, the Netherlands}

\author{Fabian Walter}
\affiliation{Max-Planck Institute for Astronomy, Konigstuhl 17, D-69117 Heidelberg, Germany}

\author{Roberto Decarli}
\affiliation{INAF – Osservatorio di Astrofisica e Scienza dello Spazio di Bologna, via Gobetti 93/3, I-40129, Bologna, Italy}

\begin{abstract}
    Feedback and outflows in galaxies that are associated with a quasar phase are expected to be pivotal in quenching the most massive galaxies. However, observations targeting the molecular outflow phase, which dominates both the mass and momentum and removes the immediate fuel for star formation, are limited in high--z QSO hosts. Massive quiescent galaxies found at $z\sim4$ are predicted to have already quenched star formation by $z\sim5$ and undergone their most intense growth at $z>6$. Here, we present two ALMA detections of molecular outflows, traced by blue--shifted absorption of the OH 119 \mum doublet, from a sample of three $z>6$ infrared luminous QSO hosts: J2310+1855 and P183+05. OH 119 \mum is also detected in emission in P183+05, and tentatively in the third source: P036+03. Using similar assumptions as for high--z Dusty Star--Forming Galaxy outflows, we find that our QSOs drive molecular outflows with comparable mass outflow rates, and that are comparably energetic except for J2310+1855's significantly lower outflow energy flux. We do not find evidence, nor require additional input from the central AGN to drive the molecular outflow in J2310+1855 but can not rule out an AGN contribution in P183+05 if a significant AGN contribution to $\rm L_{FIR}$ is assumed and/or if the outflow covering fraction is high ($\geq53\%$), which evidence from the literature suggests is unlikely in these sources. Differences observed in the blue--shifted absorption spectral properties may instead be caused by the QSO hosts' more compact dust continuum, limiting observations to lower altitude and more central regions of the outflow.
\end{abstract}

|



\section{Introduction} \label{sec:intro}
    Feedback from star--formation and black hole activity plays a critical role in the evolution of galaxies and galaxy populations throughout their lifetimes. Energy and momentum injected back into the ISM through these processes regulates galaxy growth by heating, disturbing or ejecting gas that may have fueled future star formation. The removal of low angular momentum and metal enriched material from the centers of galaxies via outflows is an essential step in our understanding and abilities to reproduce observed disky galaxy morphologies, metallicity gradients and the polluted circum--/inter galactic medium \citep{Governato2010,Somerville1999,Keres2009,Veilleux2005,MeyerYork1987,Simcoe2004,Travascio2020}. 
    
    Regulated galaxy growth implies that the most rapidly evolving galaxies must also experience the most vigorous feedback. For massive quiescent galaxies observed as early as $z>4$ \citep{Straatman2014,Guarnieri2019,Carnall2020, Valentino2020}, their short but explosive lifetimes must have been accompanied by extreme star--formation and subsequent feedback and outflows, in order to deplete their gas reserves by this time \citep{Glazebrook2017,EstradaCarpenter2020,Forrest2020}. High--z ($z>4$) dusty star--forming galaxies (DSFG) that also host an AGN thus present the most probable progenitors of such systems, warranting an investigation into the scale and impact of their outflows.
    
    At low redshift, observations of the ionised and neutral phases have shown that outflows are ubiquitous \citep{Veilleux2005,Heckman2017,Rupke2018,Veilleux2020}. With the advent of Herschel, observations extended to the cooler neutral and molecular phases and revealed that the associated denser gas that is directly associated with star formation, dominates the mass and momentum budget of their outflows \citep{Fluetsch2021}. The first such detections uncovered a molecular outflow in Mrk 231 via emission in high velocity line wings of the CO molecule \cite{Feruglio2010} and via absorption in a P--Cygni profile of the OH molecule \cite{Fischer2010}. Outflows have since been detected in various atomic and molecular species (CO, H$_2$O, HCN, [CII], [CI], $\OHp$, OH, $\CHp$, $\H2Op$, etc.), using the same techniques. At high--z, however, high velocity line wings which typically make up only a few percent of the total emission line flux, become difficult to detect and disentangle, making CO an inefficient tracer. Even observations of the bright [CII] line has detected outflows in only a handful of individual galaxies (e.g., \citealt{Fan2018,HerreraCamus2021,Tripodi2022}), and can depend strongly on the methodology implemented \cite{Maiolino2012,Cicone2015,Meyer2022}. Stacking has helped to provide additional detections \citep{Gallerani2018,Ginolfi2020}, but not without some contradictory results \citep{Bischetti2019,Novak2020}. Most convincingly, \cite{Spilker2020a} show in their Sec. 5.2 that even high quality [CII] spectra, with comparable or higher S/N than the stacked results mentioned, do not show evidence of high velocity wings in sources with known molecular outflows, thus deeming [CII] as an unreliable outflow tracer of molecular gas.
    
    Absorption lines, on the other hand, have proven to efficiently trace the gas intervening the observer and host galaxy in dusty high--z sources \citep{Falgarone2017,Indriolo2018,Spilker2020a,Spilker2020b,Berta2021,Riechers2021,Butler2021,Shao2022}, reliably detecting in-- and out--flowing gas when red-- or blue--shifted with respect to the host galaxies systemic velocity. 
    
    The OH molecule is the most extensively utilised of such species, with samples in the local universe spanning both star--forming, composite and AGN dominated systems \citep{Sturm2011,Veilleux2013,Spoon2013,Stone2016,Calderon2016,GonzalezAlfonso2017}. Consistently throughout these studies, outflow properties are found to best correlate with the AGN luminosity ${\rm L_{AGN}}$ or AGN fraction $\alpha_{\rm AGN}$ of the host galaxy, and at best only weakly with the host galaxy's star--forming properties \citep{Calderon2016}. Higher ${\rm L_{AGN}}$ and $\alpha_{\rm AGN}$ drive faster outflow velocities \citep{Sturm2011,Spoon2013,Stone2016} and shorter gas depletion time scales \citep{Sturm2011,GonzalezAlfonso2017}. Total OH absorption line EWs strongly correlate with 9.7\mum silicate absorption, a measure of obscuration within the host galaxy \citep{Stone2016}, whilst the relative strength of the OH emission component decreases \citep{Spoon2013,Veilleux2013}. Furthermore, pure OH emission spectra found in some AGN dominated systems display relatively narrow line widths, suggested to be indicative of a stage after which the AGN has cleared the obscuring material \citep{Veilleux2013}. 
    
    At high--z, observations targeting OH are mostly limited to starforming galaxies \citep{Spilker2018,Spilker2020a,Spilker2020b}, and one tentative absorption detection in the z = 6.13 QSO, ULAS J131911+095051 \citep{HerreraCamus2020}. OH absorption is detected in all of the 11 SPT sources targeted by \cite{Spilker2020a,Spilker2020b}, with no sources displaying OH in emission. At a similar rate to low--z ULIRGs and QSOs, 73\% display unambiguous evidence of outflows, none of which could be detected by high velocity wings in their corresponding [CII] spectra. Such a high detection rate indicates a typical outflow geometry that has either a wide opening angle or is widespread. In either case, the outflows display significant clumping unlike the smooth dust continuum profiles of these galaxies. Covering fractions of the clumpy outflows increase with both galaxy infrared luminosity and outflow velocity. 
    
    With no evidence and no requirement of AGN feedback needed to drive the molecular outflows studied in the high--z DSFG sample, and with only one tentative detection of OH absorption in a QSO at $z{>}6$, there is a clear gap in our understanding of how and if AGN activity contributes to the ejection of molecular outflows in the early universe. In this paper we present molecular gas observations of the 119 \mum OH doublet in three luminous $z{>}6$ QSOs and their FIR bright host galaxies: J231038.88+185519.7 (hereafter J2310+1855, z=6.00282), PSO J183.1124+05.0926 (hereafter P183+05, z=6.43862) and PSO J036.5078+03.0498 (hereafter P036+03, z=6.54052).
    
    \vspace{-26pt}
    \begin{deluxetable*}{ccccccccccc}
        \tablecaption{QSO Host Galaxy Properties From the Literature. \label{tab:Litvals}}
        \tablewidth{\textwidth}
        \tablehead{
        \colhead{Name}&\colhead{$\rm z_{[CII]}$}&\colhead{}&\colhead{$\rm L_{\rm[CII]}$}&\colhead{$\rm L_{\rm FIR}$}&\colhead{$\rm M_{1450}$}&\colhead{$\rm M_{\rm gas}$}&\colhead{$\rm M_{\rm dyn}$}&\colhead{$i$}&\colhead{$\rm v_{\rm rot}$}&\colhead{$\rm v_{\rm rot}/\sigma_{\rm V}$}\\
        \colhead{} & \colhead{}& \colhead{$\rm [kpc\ ^{\prime\prime-1}]$} & \colhead{$\rm 10^{9}[L_{\odot}]$} & \colhead{$\rm 10^{12}[L_{\odot}]$} & \colhead{[mag]} & \colhead{$\rm 10^{10}[M_{\odot}]$} & \colhead{$\rm 10^{10}[M_{\odot}]$}& \colhead{[deg]}&\colhead{[\kms]} &\colhead{[\kms]}
        }
        \startdata
        J2310+1855  & $6.00282^{\rm\ d}$ & 5.84 & $8.83\pm0.44^{\rm\ c}$ & $15.0^{+0.39\rm\ e}_{-0.33}$ & $-27.75^{\rm\ d}$ & $4.4\pm0.2^{\rm\ e}$ & $5.2^{+2.3\rm\ e}_{-3.2}$ & $25^{\rm\ e}$ & $\simeq347^{\rm\ e}$ & $\sim6^{\rm\ e}$\\
        P183+05 & $6.43862^{\rm\ b}$ & 5.61 & $7.15\pm0.32^{\rm\ b}$ & $10.53\pm0.36^{\rm\ b}$ & $-26.99^{\rm\ d}$ & $5.0^{+27.8\rm\ a}_{-2.1}$ & $>13.0^{+7.8\rm\ a}_{-9.1}$ & $<22^{\rm\ a}$ & $>320^{\rm\ a}$ & $2.29^{\rm\ a}$\\
        P036+03 & $6.54052^{\rm\ b}$ & 5.56 & $3.38\pm0.09^{\rm\ b}$ & $5.77\pm0.12^{\rm\ b}$ & $-27.28^{\rm\ d}$ & $2.8^{+15.4\rm\ a}_{-1.1}$ & $2.9^{+1.1\rm\ a}_{-0.7}$ & $21^{+5\rm\ a}_{-4}$ & $200^{+50\rm\ a}_{-30}$ & $3.21^{\rm\ a}$\\
        \enddata
        \tablecomments{Properties of the QSOs and host galaxies taken from the literature. To obtain $\rm L_{\rm FIR}$ for J2310 we find the conversion between $\rm L_{\rm TIR}$ and $\rm L_{\rm FIR}$ using the modified blackbody component of the best-fit SED of \cite{Tripodi2022}. This gives $\rm L_{\rm TIR}/L_{\rm FIR}=1.6$.  References: \textbf{a)} \cite{Neeleman2021}, \textbf{b)} \cite{Venemans2020}, \textbf{c)} \cite{Wang2013}, \textbf{d)} \cite{Mazzucchelli2017}, \textbf{e)} \cite{Tripodi2022}. 
        }
    \end{deluxetable*}
    
    In Sec. \ref{sec:obs} we present the ALMA band 7 observations, reduction and imaging of the OH 119\mum line and continuum in these sources. In Sec. \ref{sec:results} we explore the spectral properties of the OH line and relationship to host galaxy properties, making comparisons to the high--z DSFG sample of \cite{Spilker2020a,Spilker2020b} and low--z sample of \cite{Veilleux2013}. In Sec. \ref{sec:outflowprops} we derive outflow properties, consider the driving mechanisms required, as well as the possible impact on the future of the host galaxy. In Sec. \ref{sec:discussion} we discuss the the significance of spectral differences found between the QSO and high--z DSFG sample and the context of the observed molecular outflows within the evolution of the host galaxies and QSOs. In Sec. \ref{sec:conclusion} we summarise our conclusions. Throughout the paper we adopt a flat $\Lambda$CDM cosmology with $\Omega_{\rm m}= 0.307$ and ${\rm H_0 = 67.7\ km\ s^{-1}\ Mpc^{-1}}$ \citep{Planck2016}.
  
\section{Sample, Observations and Imaging} \label{sec:obs}
    The data presented in this paper were taken as part of the ALMA Cycle 6 project 2018.1.01790.S (P.I.: P.P. van der Werf) targeting 3 far--IR bright QSOs: J2310+1855, first identified in the Sloan Digital Sky Survey (SDSS) \citep{Jiang2016}, and P183+05 and P036+03, both selected as z--dropouts in the PanSTARRS1 survey \citep{Chambers2016} and first reported by \cite{Mazzucchelli2017} and \cite{Venemans2015}, respectively. \cite{Banados2019} reported the presence of a damped $\rm Ly\alpha$ absorber (DLA) in the proximity zone of P183+05, at z=6.40392. At this redshift the OH 119 \mum doublet lies just outside our frequency coverage at 340 GHz, blue--shifted $-1400\ \kms$ with respect to the systemic velocity of P183+05's host galaxy. All three sources have previously been studied in resolved FIR and [CII] emission with kinematic modelling \citep{Wang2013,Banados2015,Decarli2018,Feruglio2018,Shao2019,Pensabene2020,Venemans2020,Neeleman2021,Tripodi2022} and observed in various additional lines \citep{Venemans2015,Chambers2016,Jiang2016,Feruglio2018,Li2020,Banados2019,Tripodi2022,Shao2022,Decarli2022}. A summary of key properties taken from the literature is shown in Tab. \ref{tab:Litvals}.
    
    For each source, the ALMA band 7 receivers were tuned such that two overlapping spectral windows of one sideband covered the OH 119 $\mum$ line, with the remaining two spectral windows placed to detect the underlying dust continuum at high S/N. All sources were observed with 43 ALMA antennas, in good conditions. Specific details of the observations and imaging for each source can be found in Tab. \ref{tab:obsspecs}.
    
    The raw data was reduced using CASA \citep{McMullin2007} version 5.4.0-70 for J2310+1855 and CASA version 5.4.0-68 for P183+05 and P036+03. All sources were non-interactively imaged using \texttt{tclean} down to a cleaning threshold of $1\sigma$ and a robust weighting of 0.5, which was found to optimised the signal--to--noise without losing significant spatial resolution. We separate the data into two data cubes for each side band, with no continuum subtraction and leave the frequency resolution equal to the channel resolution of 15.624464 MHz. The rest--frame 119 \mum continuum map was created using the available line--free channels in each source.
        
    \addtolength{\tabcolsep}{-3pt}       
    \begin{deluxetable}{llccc} 
        \tablecaption{ALMA Observations\label{tab:obsspecs}}
        \tablewidth{\textwidth}
        \tablehead{\colhead{} & \colhead{} & \colhead{J2310+1855} & \colhead{P183+05} & \colhead{P036+03}}
        \startdata
        R.A.& [hh:mm:ss]  & 23:10:38.90   & 12:12:26.98  & 02:26:01.88 \\[0.5mm]
        Dec.& [dd:mm:ss]  & +18:55:19.82  & +05:05:33.49 & +03:02:59.39 \\[0.5mm]
        $\rm t_{obs}$ & [min]  & 34.50 & 44.28  & 39.50 \\[0.5mm]
        Beam & [$^{\prime\prime}$]  & $0.56\times0.48$  & $0.58\times0.45$  & $0.63\times0.44$ \\[0.5mm]
        $\sigma_{\rm cont}$ & [$\rm\mu\rm Jy\ beam^{-1}$]  &  3.31 & 0.65 & 0.99 \\[0.5mm]
        $\Delta{\rm chan}$& [\kms]  & 13.1 & 13.9 & 14.1 \\[1mm]
        $\overline\sigma_{\rm chan}$ & [$\rm\mu\rm Jy\ beam^{-1}$] & 12.5 & 4.18 & 6.48 \\[1mm]
        \enddata
        \tablecomments{$\overline\sigma_{\rm chan}$ is the mean channel sensitivity, for a channel width of $\Delta{\rm chan}$, in the sideband containing the OH 119\mum doublet.}
    \end{deluxetable}

\section{Results} \label{sec:results}
    \subsection{Spectra and Spectral Fitting} \label{subsec:specfit}
        \begin{figure*}
            \plotone{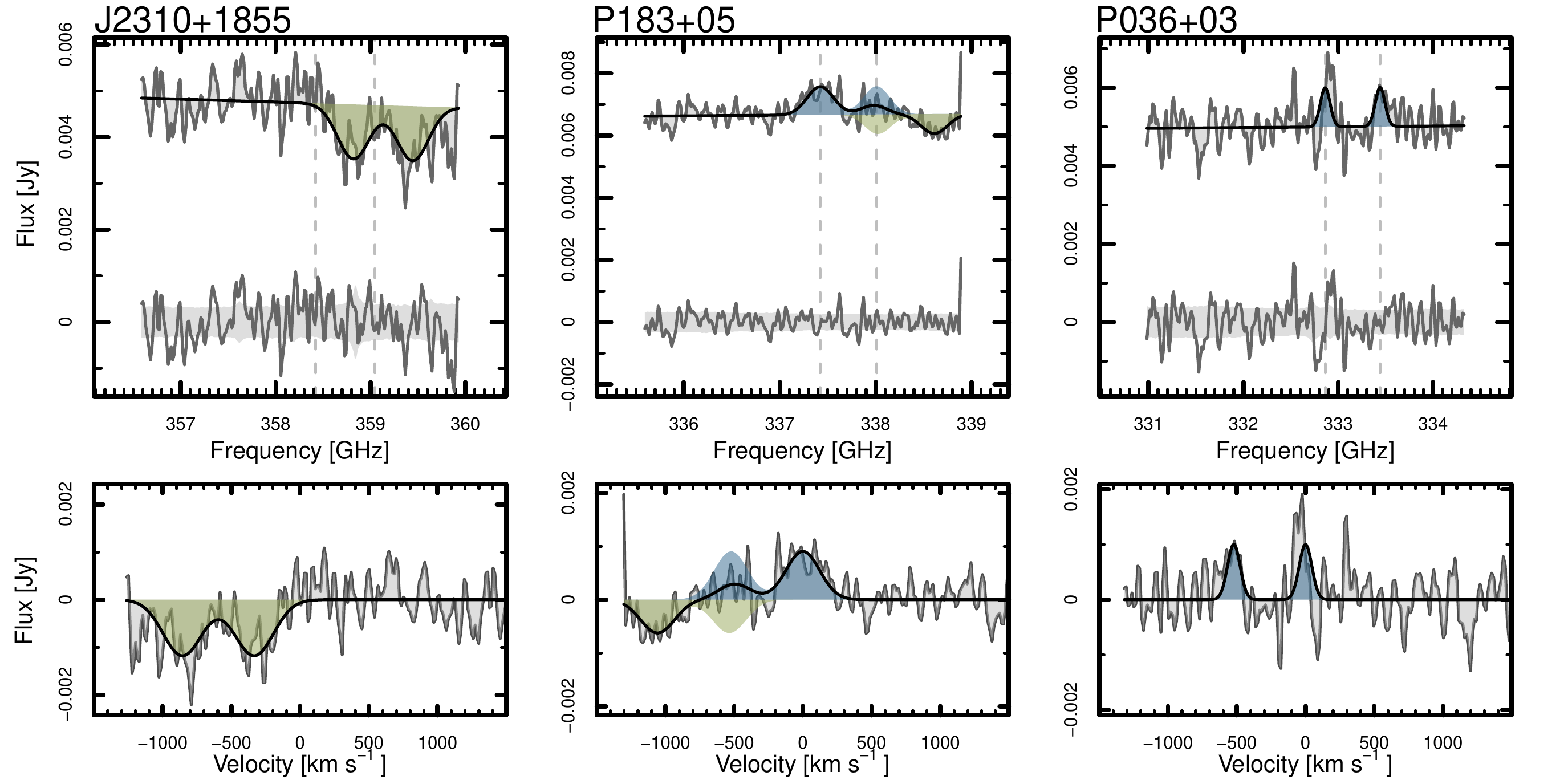}
            \caption{Spectra, with resolution of 15.6 MHz, of the three QSOs in this paper left to right: J2310+1855, P183+05 and P036+03. \textbf{Top Panels:} ALMA spectra of the side band covering the OH 119 \mum doublet, as a function of observed frequency with the underling continuum included. The best fit of each spectra are over plotted as a solid black line with individual absorption and emission components shaded in green or blue, respectively. The residuals after subtracting the best fits is shown at the bottom with the channel by channel RMS shaded in grey. The rest frequencies of the OH 119 \mum doublets, assuming the redshifts in Tab. \ref{tab:Litvals}, are shown by the vertical dashed lines. \textbf{Bottom Panels:} The same spectra with the best fit continuum subtracted, shown as a function of velocity.  \label{fig:fits}}
        \end{figure*}
        
\begin{figure}
            \plotone{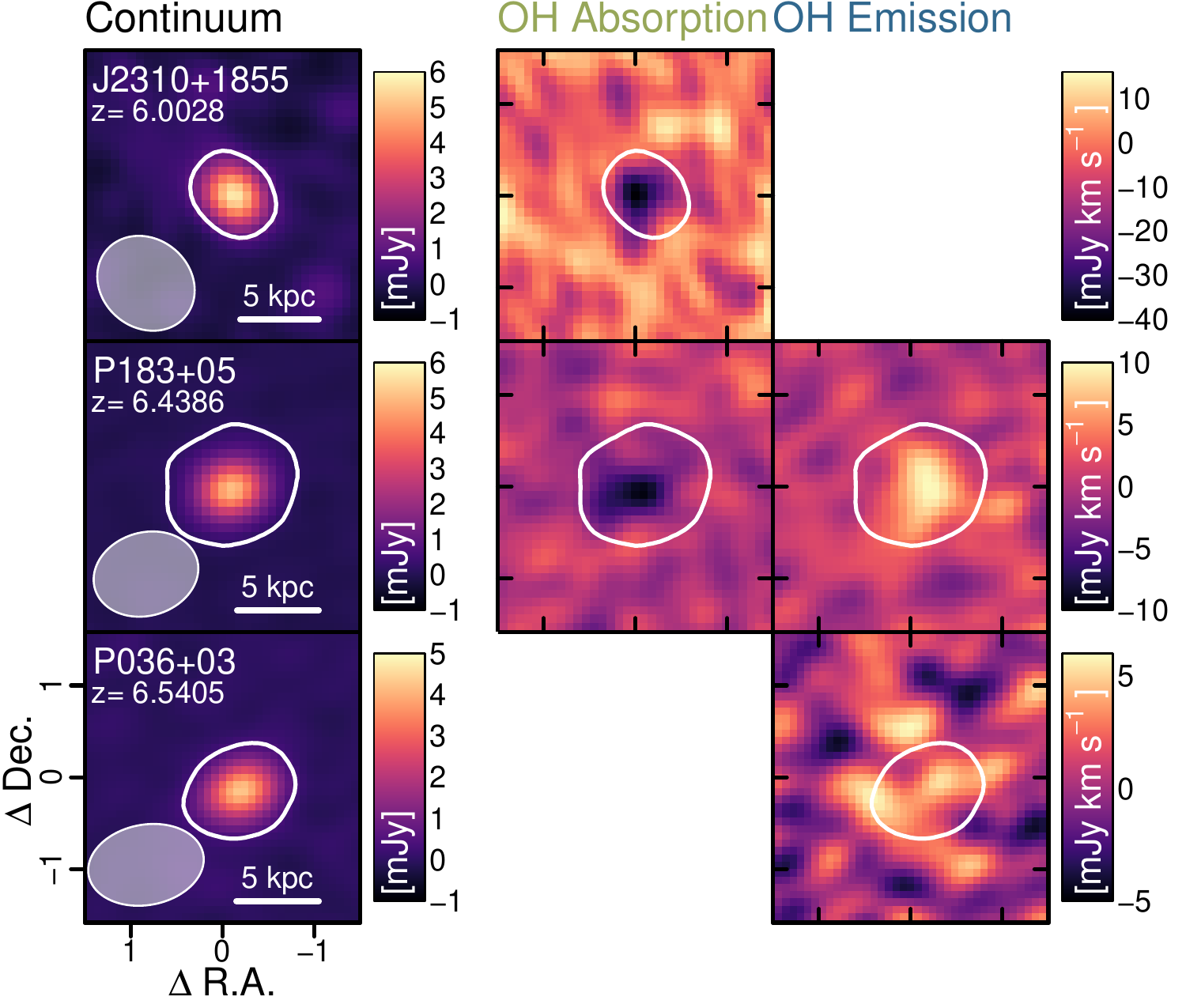}
            \caption{Maps of the three QSOs in this paper top to bottom: J2310+1855, P183+05 and P036+03. From the left: 119 \mum continuum intensity maps with the ALMA beam shown as the white ellipse in the lower left corner. This is followed by the OH 119 \mum  absorption line (J2310+1855 and P183+05) and/or OH 119 \mum emission (P183+05 and P036+03) line map. Lines are extracted from channels within the FWHM of the higher and lower frequency doublet, for the absorption and emission, respectively. The $8\sigma$ continuum level contour is overplotted in white for all maps, indicating the stacking region used to create the spectra shown in Fig.\ \ref{fig:fits}. \label{fig:Ims}}
        \end{figure}

        The rest frame 119 \mum dust continuum is detected at high S/N and marginally resolved in each of our sources (Fig.\ \ref{fig:Ims}). The integrated spectra shown in Fig.\ \ref{fig:fits} are created by stacking all spaxels with a continuum level $>8\sigma$ in the corresponding continuum map. This is done for both side bands. The OH 119\mum doublet is clearly detected in absorption in two sources (J2310+1855 and P183+05), both blue--shifted with respect to the systemic velocity of their respective host galaxies. OH 119\mum absorption is not found at systemic velocities in any of the sources. OH 119\mum is also detected in emission at systemic velocity in P183+05, forming a P--Cygni profile in the spectra of this source, and tentatively detected at systemic velocity in P036+03. This is in contrast with the sample of high--z DSFGs studied by \cite{Spilker2020a}, where OH 119\mum emission was not detected in any of the sources. We discuss this result in Sec. \ref{subsec:OHemission}.
        
        We fit each spectrum across both side bands, with a combination of one or two double--Gaussians and an underlying linear dust continuum slope. The corresponding spectral ranges used are 344.0-360.0 GHz, 335.5-351.0 GHz and 330.9-346.5 GHz for J2310+1855, P183+05 and P036+03, respectively. The double--Gaussian component consists of two Gaussians of equal amplitude and width, and placed at a fixed rest--frame separation of 4.37 GHz, as required of the OH $\Lambda$--doublet at the redshift under consideration. The continuum flux density and slope are left as free parameters as well as the amplitude and width of the double--Gaussian. The frequency offset of the absorption features in J2310+1855 and P183+05 is left as a free parameter whilst the OH emission features in P183+05 and P036+03 are fixed at the systemic velocity of their hosts. The resulting spectral fits are shown in Fig.\ \ref{fig:fits} along with the residual spectra. The residuals of J2310+1855 display possible excess absorption at more blue--shifted velocities than is captured by the single double--Gaussian fit, possibly indicating an outflow at even higher velocities. This feature lies at the edge of the side band, however, and additional observations extending to higher frequencies are needed to confirm an additional outflow component. The narrow ($108\pm22\ \kms$) line width of the tentative OH emission in P036 is comparable to that found for the weak water vapour emission of the blended $3_{22}-3_{13}$ and $5_{23}-4_{32}$ $\rm H_2O$ transitions \citep{Banados2015}, which similarly traces warm, dense molecular gas. The best fit parameter values and uncertainties for each source are presented in Tab. \ref{tab:specfitpars}. Line maps, integrated over the FWHM, are shown in Fig. \ref{fig:Ims}. 
        
\begin{figure*}
            \includegraphics[width=470pt]{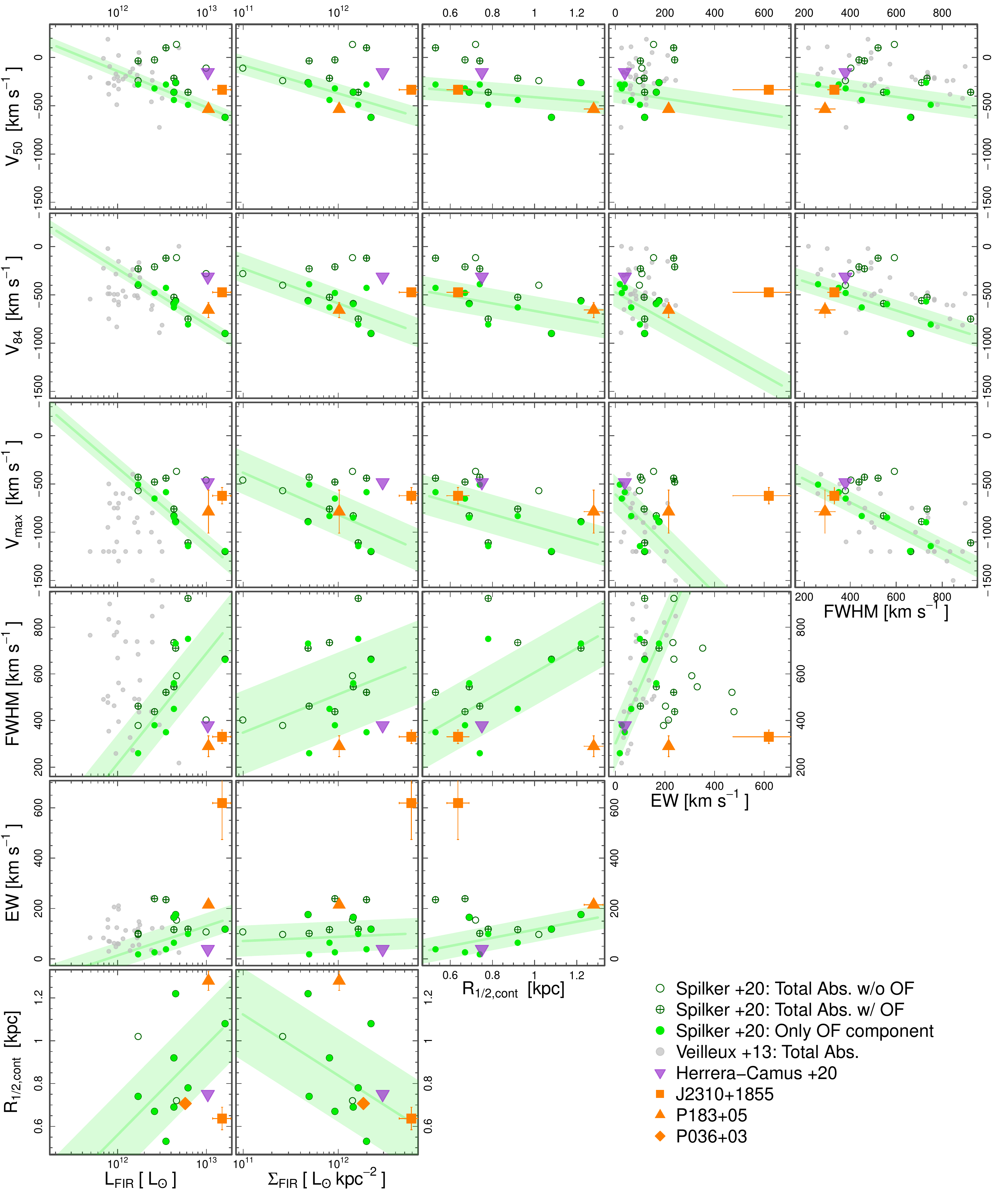}
            \caption{Grid of host galaxy and best fit OH 119 \mum absorption spectra properties. This includes: $\mathbf{v_{50}}$: mean velocity, $\mathbf{v_{84}}$: velocity above which 84\% of the absorption lies, $\mathbf{v_{max}}$: velocity above which 98\% of the absorption lies, \textbf{FWHM:} full width half maximum of the absorption, \textbf{EW:} equivalent width of the absorption, $\mathbf{L_{IR}}$: Far Infrared luminosity of the host galaxy, $\mathbf{\Sigma_{FIR}}$: Infrared surface brightness of the host galaxy given by $\rm L_{FIR}/(2\pi R_{1/2,cont}^2)$, $\mathbf{R_{cont}}$: Effective radius of the 119\mum continuum emission. We show low--z sources in grey, including both star forming and AGN host galaxies \citep{Veilleux2013}. The spectral properties are measurements of the total (systemic and blue shifted) absorption lines. In green we show the high--z DSFG sample studied by \cite{Spilker2020a,Spilker2020b}, further separating this sample into measurements of the total (systemic and blue shifted) absorption line (dark green crossed--circles), just the systemic components (dark green open circles) and just the outflowing components (light green filled). Sources with only an outflowing component will appear as a light green dot with a dark green outline. We display a fit to the outflow--only components in each panel (illustrative only) with a solid line and shade the $1\sigma$ scatter, both in light green. We include the tentative OH detection in the z = 6.13 QSO, ULAS J131911+095051 (\citealt{HerreraCamus2020}, purple triangle). J2310+1855 and P183+05 are shown by the filled orange square and triangle, respectively.). \label{fig:ParGrid}}
        \end{figure*}

        As is commonly done in the literature, we derive additional quantities of the absorption lines from our best fit parameters in order to compare our work with previous OH studies \citep{Veilleux2013,Spilker2020a,HerreraCamus2020}. These include the full width half maximum, equivalent width (EW, equal to optical depth in the optically thin regime) and three definitions of the absorption line velocity: $v_{50}$, $v_{84}$ and $v_{max}$, referring to the velocities above which 50\%, 84\% and 98\% of the absorption lies, respectively. These values are also found in Tab. \ref{tab:specfitpars} and presented in Fig.\ \ref{fig:ParGrid} as functions of FIR luminosity and EW.
        
        For comparison, Fig.\ \ref{fig:ParGrid} also shows results from previous samples of OH absorption in star--forming and AGN hosts at both low--z \citep{Veilleux2013,Stone2016}, and high--z \citep{Spilker2020a,HerreraCamus2020}. As we are specifically interested in the outflowing molecular gas we additionally plot the outflow--only components of the high--z DSFGs studied by \citet{Spilker2020a} (the separate components are not provided for the low--z sample). The outflow--only components in general show stronger correlations than the full absorption line values. Furthermore, we find that the strongest correlation with outflow velocity is associated with $V_{50}$ and not $V_{max}$, as is found in studies that use the total (systemic and blue--shifted) absorption lines. The tighter relation found with $V_{max}$ for the full absorption line may be due to the increasing fractional contribution of the outflowing gas to the total absorption signature towards more negative velocities of the absorption line. We provide a fit to the outflow--only components of the DSFG sample in each panel of Fig.\ \ref{fig:ParGrid} to guide the reader's eye when comparing these values with the OH outflows in J2310+1855 and P183+05. These fits are illustrative only and are not intended to imply true physical trends in all panels. 
        
        A notable difference evident between the OH outflows in the DSFGs and those of J2310+1855 and P183+05 is the offset to smaller FWHMs at similar $\rm L_{FIR}$ for the two QSOs. The central outflow velocities ($v_{50}$) of the two QSOs appear on or near the DSFG $v_{50}$ vs $L_{FIR}$ trend. The narrower line widths of the QSO outflow lines, however, cause the QSO outflows to drift towards slower than expected velocities for more extreme velocity definitions ($v_{84}$ and $v_{max}$ vs $L_{FIR}$). This is also seen in the plots of EW vs V. 
        
        A second, notable observation is the large EW of the OH absorption in J2310+1855. Although less extreme in the case of P183+05, the EWs of the OH absorption lines are relatively larger in both J2310+1855 and P183+05, than seen in the DSFGs, given the narrower line widths. 
        
        Care should be taken when interpreting the position of ULAS J131911+095051 \citep{HerreraCamus2020} in Fig.\ \ref{fig:ParGrid} as these values are taken from a single Gaussian fit. The lower frequency transition of the OH 119 \mum doublet is likely cancelled out by emission at systemic velocities, as is seen in P183+05. Some absorption from the higher frequency transition is also likely lost on the red side of the line, meaning a fit to this absorption feature alone will underestimate the absorption strength and line width and will overestimate the blue--shifted velocity offset. 
        
    \subsection{Source Sizes}
        \begin{figure*}[b]
            \gridline{\fig{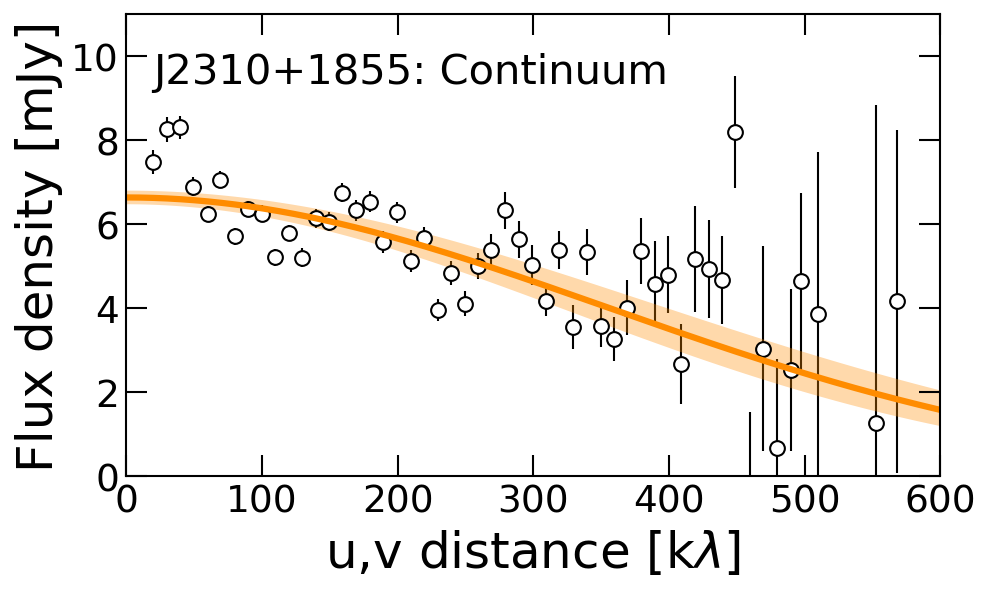}{0.3\textwidth}{}
                      \fig{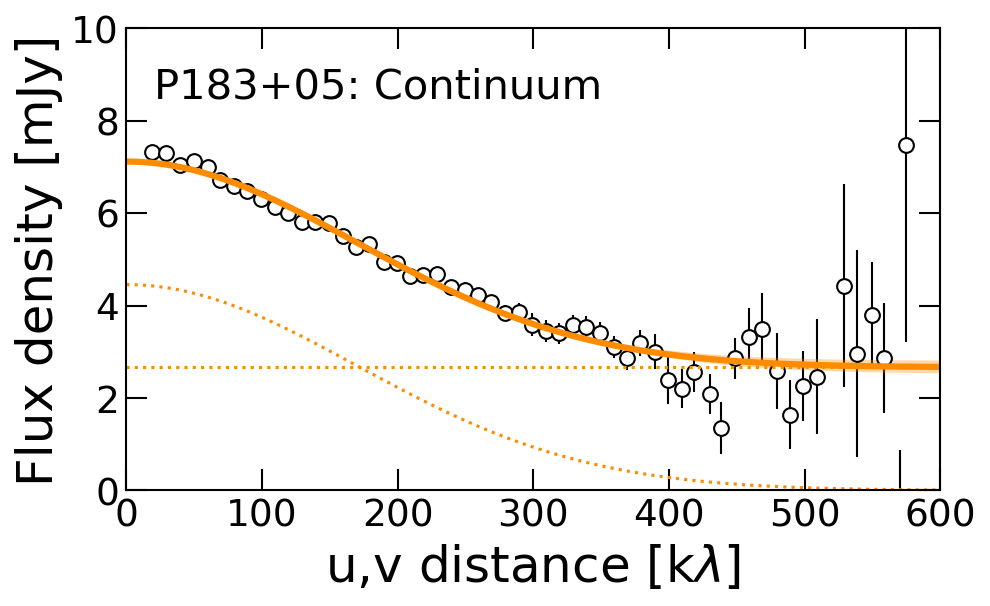}{0.3\textwidth}{}
                      \fig{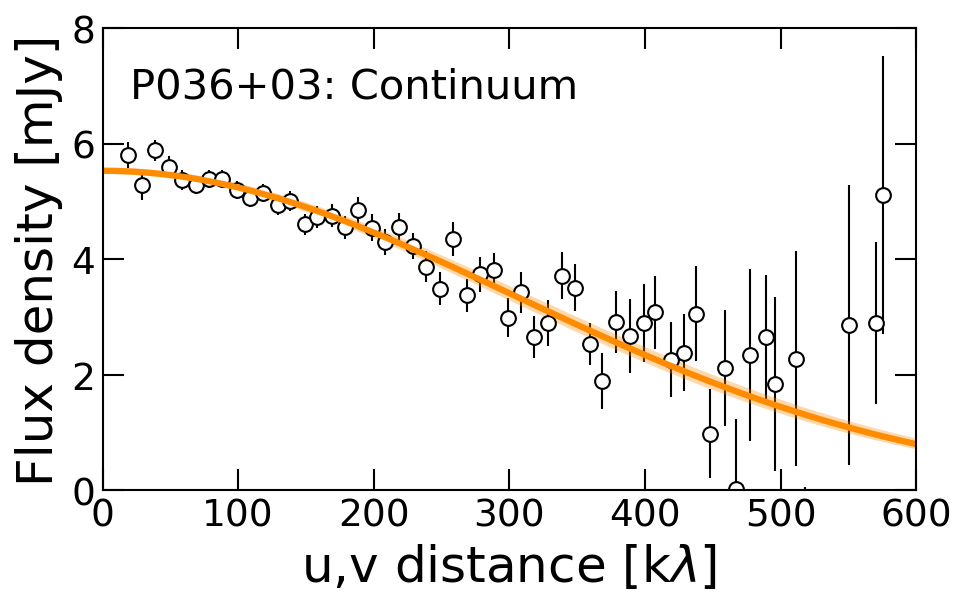}{0.3\textwidth}{}
                      }
            \gridline{\fig{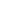}{0.3\textwidth}{}
                      \fig{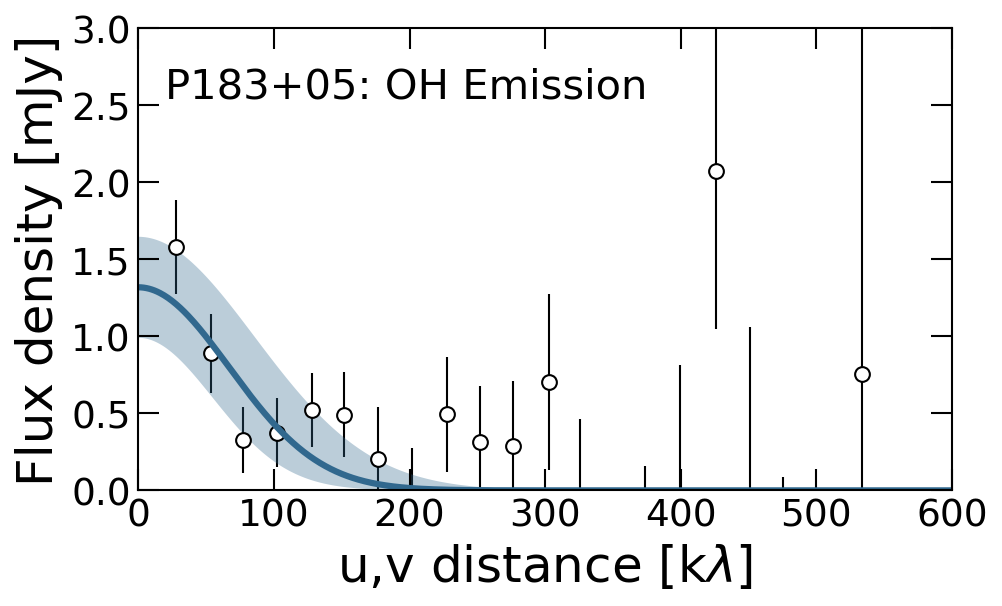}{0.3\textwidth}{}
                      \fig{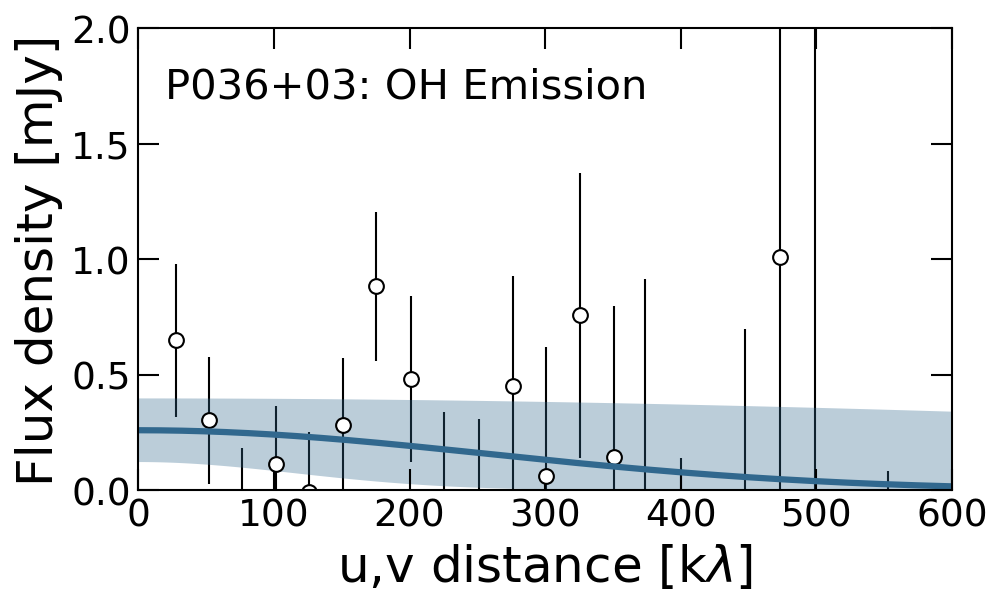}{0.3\textwidth}{}
                      }
            \caption{Radial profiles of the dust continuum (top row) and OH 119 \mum emission (bottom row) visibilities, using UV bins of width 10 $k\lambda$. Mean values of the real visibilities in each bin are taken to fit a Gaussian or combination of a Gaussian and point source profile to each source, which we over plot in orange/blue along with with the $1\sigma$ uncertainty shaded also in orange/blue. We show the individual components of the combination fit to the continuum of P183 with dotted orange curves. \label{fig:sizes}}
        \end{figure*}
    
        For each source we estimate the size of the 119 µm continuum emitting area. Taking only line–free channels, the visibilities of each data set are binned in radial uv bins. We perform the procedure for bins of uv widths of 10, 25 and 50 $k\lambda$ taking the mean value of the real visibilities in each bin. We perform the fit on the resulting binned data using the scipy’s \texttt{curve\_fit} routine, adopting a scatter in the visibilities as the uncertainty. We fit the continuum data with (1) a single Gaussian; (2) a combination of Gaussian and a point source. A single Gaussian fit is preferred for all the sources except P183+05. We find mean 119 µm continuum half–light radii of $0.109\pm0.009^{\prime\prime}=637$ pc, $0.228\pm0.008^{\prime\prime}=1280$ pc and $0.127\pm0.004^{\prime\prime}=707$ pc, for J2310+1855, P183+05 and P036+03, respectively. The errors include the formal uncertainties from \texttt{curve\_fit} and the scatter between different uv-binnings. The results are robust with respect to the uv bin size. All sources are marginally resolved by the beam as seen by the distinct drop--off in continuum flux towards larger uv distances in each source (Fig.\ \ref{fig:sizes}).

        Our measured half--light radii of J2310+1855 and P183+05 are comparable to previous high resolution measurements of the dust continuum at 260 GHz ($\rm R_{1/2}=0.11^{\prime\prime}\times0.095^{\prime\prime}$, \citealt{Tripodi2022}) and 158 \mum($0.225^{\prime\prime}\times0.175^{\prime\prime}$ \citealt{Venemans2020}), respectively. P036+03's measured half--light radius is large compared to previous high resolution measurements of $0.095^{\prime\prime}\times0.08^{\prime\prime}$ at 158 \mum \citep{Venemans2020}. All three QSO sizes are within the range reported for the full 27 QSO hosts studied by \cite{Venemans2020} and the high--z comparison sample of DSFGs \citep{Spilker2020b}.
        
        The same procedure is also applied to continuum subtracted channels within the FWHM of the lower--frequency line of the OH emission doublet in P183+05, and in both emission components of P036+03 (bottom row of Fig.\ \ref{fig:sizes}). This provides an OH 119 \mum emitting radius of $0.63\pm0.13^{\prime\prime}=3.54$ kpc for P183+05. The OH emission is not resolved in the case of P036+03. 
        
    \subsection{Outflow Covering Fractions and Detection Rates}\label{subsec:fcov}
        A limiting factor of absorption line work is that only gas lying between the observer and the background continuum may be detected. For some geometries, outflowing or inflowing gas may or may not intervene the line of sight (LOS) and background continuum given particular viewing angles. Consequently, a non--detection does not immediately imply the absence of an out/inflow. Likewise, when a detection is made, the observed covering fraction $\rm f_{\rm cov}$ of the intervening gas (the fraction of the background continuum with an absorption signature), may not be indicative of the overall $4\pi$ $\rm f_{\rm cov}$, unless the geometry is known to be isotropic. Assumptions must therefore be made when converting observed column densities to total, galaxy--wide masses (see Sec. \ref{sec:outflowprops}).
        
        Measuring the LOS covering fraction requires resolved observations of the intervening gas. Our observations do not resolve the structure of the outflows detected in OH absorption in J2310+1855 and P183+05. We can, however, derive a hard lower bound on the LOS $\rm f_{\rm cov}$ by assuming an optically thick outflow and taking the fractional absorption depth of the OH line. This gives $\rm f_{\rm cov}\geq$25\% and 9.4\% for J2310+1855 and P183+05, respectively, which are comparable to the 5-20\% fractional depths measured by \cite{Spilker2020a} for the high--z DSFGs sample. 
      
        The spatial resolution achieved by \cite{Spilker2020a} allowed them to detect distinct clumps in the outflowing OH absorption and measure LOS $\rm f_{\rm cov}$ between 30-85\% in their lensing reconstructions. They note that these values are upper limits as the individual clumps remain unresolved, accounting for (at least partially) the discrepancy between LOS $\rm f_{\rm cov}$ and the aforementioned fractional absorption depths. They argue that the true covering fractions are likely closer to those estimated from the peak absorption depth as OH is expected to be highly optically thick. This has similarly been assumed in low red--shift studies (e.g., \citealt{GonzalezAlfonso2017}).
        
        In any case, $\rm f_{\rm cov}$ measured by \cite{Spilker2020a} is typically smaller than the outflow detection rate of 73\% in their sample, interpreted by the authors to imply that the ouflowing gas traced by OH absorption must exit the galaxies in some `fortuitous geometry', one that does not require a special viewing angle. \cite{Spilker2020a} also find positive trends between covering fraction, FIR luminosity and outflow velocity, further suggesting that more luminous galaxies, capable of driving more extreme outflows, are either more likely to drive an outflow or drive more widespread outflows (i.e. wider opening angle/more clumps).
        
        The detection rate of OH outflows in the four currently observed QSOs at $z>6$ is either 50\% or 75\%, depending on whether we include the tentative outflow detection in the z = 6.13 QSO ULAS J131911+095051 \citep{HerreraCamus2020}. In the following section we use the fractional absorption depth of the OH line to derive lower limits of the outflow properties. As we can not measure an upper limit of the covering fraction as \cite{Spilker2020a} did with their high spatial resolution data, we simply use an upper limit of $\rm f_{\rm cov}$ of 100\%. We caution the reader when interpreting these upper limits, that a 100\% covering fraction is likely unphysical, given the evidence from the literature that OH traces optically thick, widespread and clumpy outflows, thus resulting in covering fractions more comparable with that of the fractional absorption depth.
        
\section{Derived Outflow Properties} \label{sec:outflowprops}
    In this section we derive physical and energetic properties of the molecular outflows observed in J2310+1855 and P183+05. To compare our outflows with those found in the sample of high--z DSFGs studied by \cite{Spilker2020a,Spilker2020b}, we recalculate the outflow properties for this sample using the same methods and assumptions adopted for J2310+1855 and P183+05. We consider only the blue--shifted outflowing Gaussian component of the OH absorption spectra of the DSFGs when deriving properties of the outflow and do not include the systemic gas component. This differs from the approach of \cite{Spilker2020b}, where the outflowing component was defined as absorption seen at velocities $<-200\ \kms$ in the total (systemic and outflowing) absorption line. 
    
    \subsection{Outflow Mass, Mass Outflow Rate and Depletion Times}
        In the optically thin regime, the EW of an absorption line is equal to the integrated optical depth. In this scenario, the EW can directly replace the optical depth integrated over velocity, $\int\tau_\nu\rm dv$, when deriving the column density of the OH119 \mum, given by,
        
        \begin{equation}
            \rm N_{OH} = \int\tau_\nu dv\frac{\nu^3g_l8\pi}{c^3A_{ul}g_u},
        \end{equation}
        
        where $\nu$ is the rest frequency of the line, $\rm g_u=6$ and $\rm g_l=4$ are the upper and lower statistical weights and $\rm A_{ul}=10^{-0.86618}\ s^{-1}$ is the Einstein coefficient of the transition. $\rm N_{OH}$ is then converted to a total molecular hydrogen column density assuming a conversion factor: here we assume a $\rm N_{OH}/N_{H_2}=1\times10^{-7}$ from \cite{Nguyen2018}.
        
        To convert a column density to a full outflow mass and mass outflow rate, a geometry of the outflow must first be assumed. Our marginally resolved observations do not allow a geometry of the outflows observed in J2310+1855 and P183+05 to be determined. We therefore assume a simple spherical thin shell outflow geometry, commonly used elsewhere in the literature (e.g., \citealt{Rupke2005}), which provides a conservative estimate of the mass and mass outflow rate:
        
        \begin{eqnarray}
    			{\rm\dot{M}_{OF}} &=& {\rm4\pi R_{OF}^2 \, \mu \, m_{H_2}\,f_{cov}\,N_{H_2}\frac{V_{OF}}{R_{OF}}} = {\rm\frac{M_{OF}V_{OF}}{R_{OF}}}
    	\end{eqnarray}
    	
        where $\mu=1.36$ accounts for the universal helium fraction, R is the outflow radius, assumed to be $R_{cont}$, V is the outflow velocity $V_{50}$ and $f_{cov}$ is the covering fraction of the outflowing gas.
        
        As discussed in Sec. \ref{subsec:fcov}, we can not measure $f_{cov}$ with our spatially unresolved data. Thus, for the purpose of simply comparing our two QSOs and with the sample of DSFGs, we derive all values of $\rm\dot{M}_{OF}$ using $f_{cov}=1$, providing upper limits on $\rm{M}_{OF}$ and $\rm\dot{M}_{OF}$ for the given geometry. For J2310+1855 this gives $\rm{M}_{OF}<99\times10^{8} M_\odot$ and $\rm\dot{M}_{OF}<5300\ M_\odot yr^{-1}$ and for P183+05  $\rm{M}_{OF}<140\times10^{8} M_\odot$ and $\rm\dot{M}_{OF}<5900\ M_\odot yr^{-1}$.
        
        \begin{figure}
            \includegraphics[width=\linewidth]{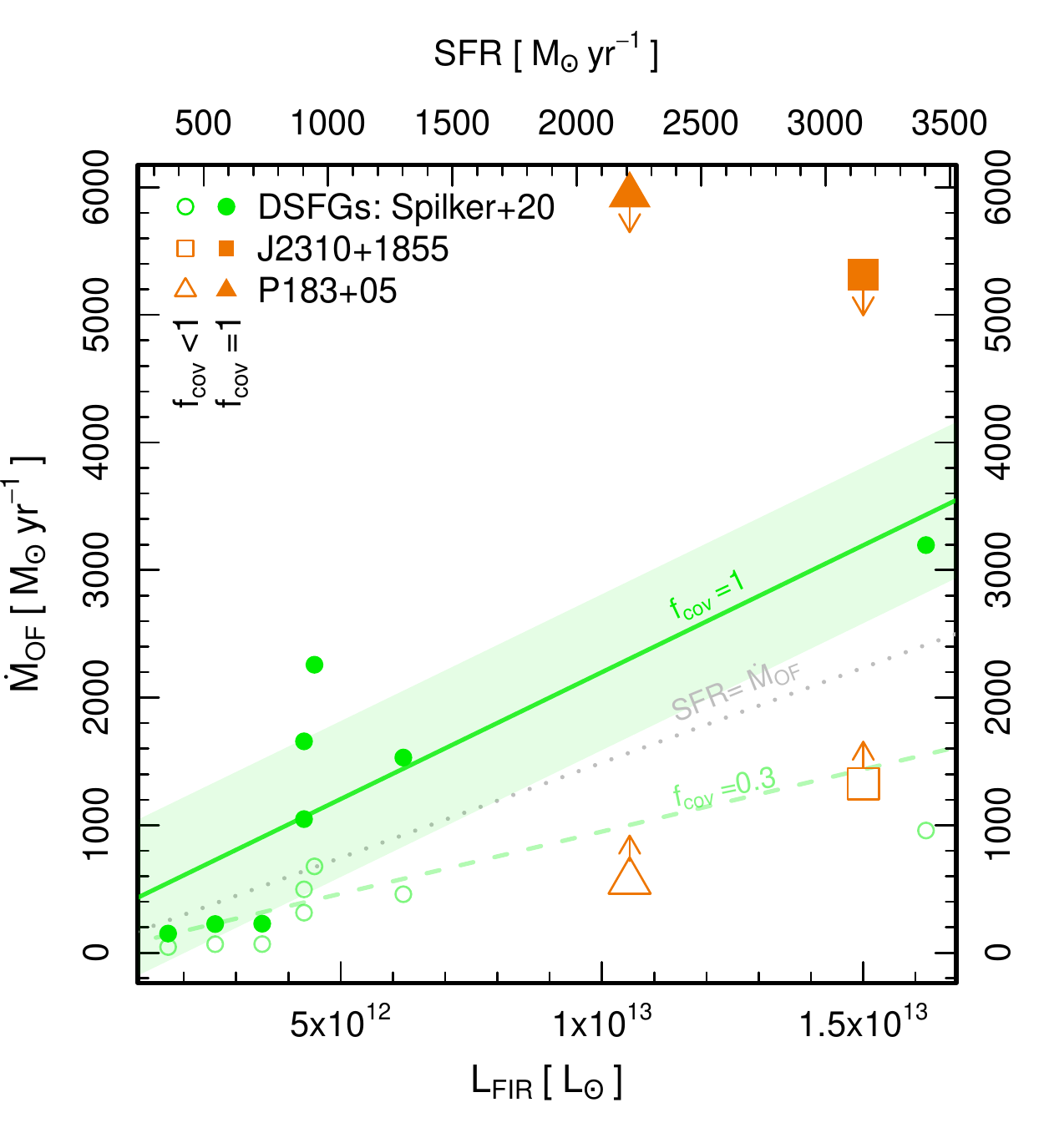}
            \caption{Mass outflow rate as a function of far infrared luminosity (bottom axis) and star formation rate (top axis), assuming a spherical thin shell geometry for all sources. We show the DSFGs \citep{Spilker2020a,Spilker2020b} in green assuming a covering fraction of 1 (filled) as upper limits and 0.3 (hollow) based on the average covering fractions measured for low--z star forming--galaxies \citep{GonzalezAlfonso2017}. We show a fit to the $f_{cov}=1$ points with the solid green line and its $1\sigma$ spread with a shade green region. A fit to the $f_{cov}=0.3$ points is shown by the dashed green line. We indicate J2310+1855 and P183+05 with an orange square and triangle, assuming a covering fraction of 1 (filled) as upper limits and at the level of their fractional OH absorption depths, 25\% and 9.4\% (hollow), respectively. The SFR=$\rm\dot{M}_{OF}$ line is shown by the grey dotted line. \label{fig:FIRvsMOFR}}
        \end{figure}
        
        Fig.\ \ref{fig:FIRvsMOFR} presents the derived values of $\rm\dot{M}_{OF}$ as a function of $\rm L_{FIR}$ (SFR). As previously reported by \cite{Spilker2020b}, the DSFGs with larger $\rm L_{FIR}$ drive outflows with higher $\rm\dot{M}_{OF}$. Focusing on the upper limit relations, we find J2310+1855 and P183+05 both sit above the DSFG relation. We note however that there is only one DSFG in the $\rm L_{FIR}$ range of the two QSO hosts, and therefore we do not view the QSO offsets to higher $\rm\dot{M}_{OF}$ as significant. In fact if the highest $\rm L_{FIR}$ DSFG is removed then a fit through the remaining DSFGs will extrapolate between the two QSOs. With the assumed geometry and a $f_{cov}=1$, we derive $\rm\dot{M}_{OF}$s greater than the SFRs in both QSOs, giving lower limits on the gas depletion time of $\rm\tau_{OF}>8.3\ Myr$ and $\rm\tau_{OF}>8.4\ Myr$ for J2310+1855 and P183+05, respectively. If star formation is included and assumed to remain constant, the combined depletion times reduce to $\rm\tau_{OF+SFR}>4.9\ Myr$ and $\rm\tau_{OF+SFR}>6.1\ Myr$ for J2310+1855 and P183+05, respectively. 
        
        \cite{GonzalezAlfonso2017} found an average $f_{cov}$ for OH absorption in their low--z star--forming galaxies of 0.3, which if adopted here would shift all but one of the DSFG and both QSO $\rm\dot{M}_{OF}$s below the $\rm\dot{M}_{OF}=SFR$ line. Whether $f_{cov}=0.3$ is a suitable covering fraction for high--z galaxies and moreover high--z QSOs, is still yet to be shown. As mentioned in Sec. \ref{subsec:fcov}, however, the fractional absorption depths of the OH absorption lines provides a hard lower limit of $f_{cov}$, which in turn can be used to derive a lower limit of $\rm{M}_{OF}$ and $\rm\dot{M}_{OF}$. This gives $\rm{M}_{OF}>25\times10^{8}\ M_\odot$ and $\rm\dot{M}_{OF}>1300\ M_\odot yr^{-1}$ for J2310+1855 and $\rm{M}_{OF}>13\times10^{8}\ M_\odot$ and $\rm\dot{M}_{OF}>560\ M_\odot yr^{-1}$ for P183+05 This corresponds to outflow depletion timescales of $\rm\tau_{OF}<33\ Myr$ and $\rm\tau_{OF}<90\ Myr$ and combined timescales of $\rm\tau_{OF+SFR}<8.8\ Myr$ and $\rm\tau_{OF+SFR}<18\ Myr$ for J2310+1855 and P183+05, respectively.
         
    \subsection{Outflow Energetics}
        The energy ${\rm\dot{E}_{OF}}$ and momentum ${\rm\dot{p}_{OF}}$ flux of an outflow can provide insight into the mechanisms needed to drive it. We derive both these values, for J2310+1855, P183+05 and the DSFGs with the same size and geometrical assumptions used to derive $\rm\dot{M}_{OF}$, giving
        
        \begin{eqnarray}
            {\rm\dot{E}_{OF}} &=& {\rm\frac{1}{2}\dot{M}_{OF}V_{OF}^2}\\
    		{\rm\dot{p}_{OF}} &=& {\rm\dot{M}_{OF}V_{OF}}.
    	\end{eqnarray}
    	
\begin{figure}
            \includegraphics[width=\linewidth]{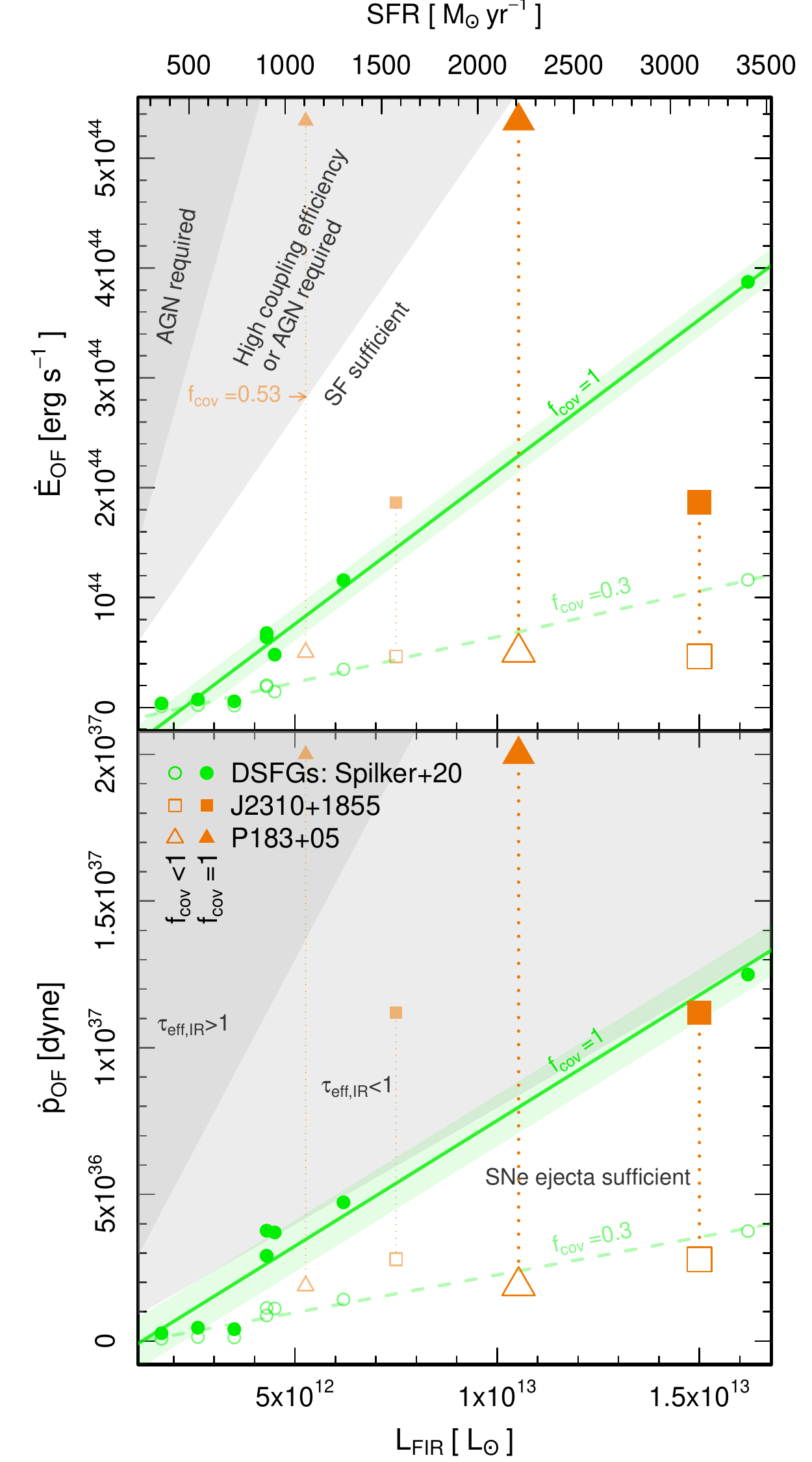}
            \caption{Energy (top) and momentum (bottom) flux as a function of $\rm L_{FIR}$ (bottom axis) and SFR (top axis) derived for the comparison sample of high--z DSFGs \citep{Spilker2020a,Spilker2020b} (green) and J2310+1855 and P183+05 (orange squares and triangles, respectively). Filled symbols indicate upper limits, assuming an outflow covering fraction of 1. Hollow symbols indicate lower limits for J2310+1855 and P183+05, assuming a covering fraction equal to the fractional OH absorption depth (25\% and 9.4\%, respectively), and values derived assuming an average covering fraction of 0.3 \citep{GonzalezAlfonso2017}, for the DSFG sample. The smaller faint orange symbols indicate ranges for J2310+1855 and P183+05 assuming a 50\% AGN contribution to $\rm L_{FIR}$ (see text). We indicate regions in the top panel where the energy injected via star formation, assuming a maximum coupling fraction to the ISM of 40\% \citep{Sharma2014,Fielding2018}, is equal to the energy flux of the outflow (white), where an unusually high coupling fraction or AGN contribution is required (light grey), and where an AGN is definitely required (darker grey). We indicate regions in the bottom panel where SNe ejecta provide sufficient momentum (white) and where radiation pressure on dust grains in an optically thin (light grey) or thick (darker grey) outflow is needed. \label{fig:FIRvsE}}
        \end{figure}

    	As in the last section, we derive these values assuming a covering fraction of 1 for all objects to compare between the DSFGs and the QSOs and again with $\rm f_{\rm cov}=0.3$, for the DSFGs, and with the fractional absorption depths of the two QSOs as lower limits (Fig \ref{fig:FIRvsE}). This gives limits of the energy fluxes of $\rm 4.7\times10^{43}\ erg s ^{-1}<{\rm\dot{E}_{\rm OF}}<19\times10^{43}\ erg s ^{-1}$ and $\rm 5.0\times10^{44}\ erg s ^{-1}<{\rm\dot{E}_{OF}}<530\times10^{44}\ erg s ^{-1}$, and of the momentum fluxes of $\rm 28\times10^{36}\ dyne<{\rm\dot{p}_{OF}}<110\times10^{36}\ dyne$ and $\rm 19\times10^{36}\ dyne<{\rm\dot{p}_{OF}}<200\times10^{36}\ dyne$, for J2310+1855 and P183+05, respectively.
    	
    	Again focusing on the upper limit relations: the DSFG sample displays a clear positive trend for both fluxes with $L_{IR}$. The QSOs sit below and above these relations for J2310+1855 and P183+05, respectively. J2310+1855's deviation below the $\rm L_{FIR}$-${\rm\dot{E}_{OF}}$ relation, however, is particularly significant. For all sources, the energy and momentum injected by star formation is sufficient to drive the observed outflow, even at the upper limit values. This is perhaps surprising for J2310+1855 and P183+05, where feedback from both star formation and the central AGN is available and may have been expected to collectively drive a stronger outflow than star formation alone. In Sec. \ref{sec:discussion} we discuss possible explanations for why we do not see evidence of an AGN contribution and instead see the opposite in the case of J2310+1855.

    \subsection{Escape Fractions} \label{sec:escfrac}
        Not all material ejected via an outflow will escape the host galaxy or the dark matter halo's gravitational potential. Instead some fraction of the material will remain bound to the galaxy and may re--accrete via a galactic fountain at a later time, unless further heating or acceleration is experienced. Some fraction of the outflow, however, may be sufficiently accelerated to pollute the CGM or even IGM, removing mass, momentum and metals from the galaxy altogether. Below we attempt to estimate a possible escape fraction of the molecular outflow in our two QSOs.
    
        Assuming that the outflow is at a radius of $\rm R_{cont}$ and that the outflow will not be decelerated by swept up material, or accelerated by radiation pressure, the dynamical masses (Table \ref{tab:Litvals}) are enough to retain effectively all of the outflowing molecular gas in both J2310+1855 and P183+05.
           
        \addtolength{\tabcolsep}{-2.5pt}   
        \begin{deluxetable}{llccc} 
            \tablecaption{Measured and Derived Host Galaxy and Outflow Properties\label{tab:specfitpars}}
            \tablewidth{\textwidth}
            \tablehead{
            \colhead{} & \colhead{} & \colhead{J2310+1855} & \colhead{P183+05} & \colhead{P036+03}
            }
            \startdata
            \multicolumn{5}{c}{119\ \mum Contintuum}\\
            $\rm R_{\rm1/2,119\ \mu m}$& [pc] & $637\pm53$  & $1280\pm45$  & $707\pm22$ \\[0.5mm]
            $\rm S_{\rm119\ \mu m}$       & [mJy] & $4.73\pm0.0016\ $ & $6.66\pm0.0010\ $ & $5.00\pm0.0020$ \\[0.5mm]
            \hline
            \multicolumn{5}{c}{OH 119\ \mum Emission (systemic)}\\
            $\rm S_{\rm OH\ 119\ \mu m}$  & [Jy \kms]  & --   & $0.28\pm0.028$ & $0.12\pm0.022$ \\[0.5mm]
            $\sigma_{\rm Emis.}$      & [\kms]& --  & $123\pm13$  & $45.9\pm9.5$ \\[0.5mm]
            $\rm FWHM_{\rm Emis}$         & [\kms]& --  & $289\pm31$  & $108\pm22$ \\[0.5mm]
            \hline
            \multicolumn{5}{c}{OH 119\ \mum Absorption (outflow)}\\
            $\rm S_{\rm OH\ 119\ \mu m}$  & [Jy \kms] & $-0.42\pm0.04$ & $-0.19\pm0.03$ & -- \\[0.5mm]
            $\rm N_{\rm OH\ 119\ \mu m}$  & $[10^{15}\ \rm cm^{-2}]$ & $8.90\pm2.1$ & $3.10\pm0.059$ & -- \\[0.5mm]
            $\rm EW_{\rm OF}$             & [\kms]& $619\pm140$  & $215\pm4.1$  & -- \\[0.5mm]
            $\sigma_{\rm OF}$             & [\kms]& $140\pm12$  & $123\pm19$  & -- \\[0.5mm]
            $\rm FWHM_{\rm OF}$           & [\kms]& $330\pm29$  & $290\pm45$  & -- \\[0.5mm]
            $\rm V_{\rm 50}$              & [\kms]& $-334\pm14$ & $-534\pm18$ & -- \\[0.5mm]
            $\rm V_{\rm 84}$              & [\kms]& $-473\pm21$ & $-656\pm79$ & -- \\[0.5mm]
            $\rm V_{\rm max}$             & [\kms]& $-622\pm85$ & $-787\pm222$ & -- \\[0.5mm]
            $\rm M_{\rm OF}$              & $[10^8\ \rm M_\odot]$ & $25-99$ & $13-140$ & -- \\[0.5mm]
            $\rm\dot{\rm M}_{\rm OF}$ & $[\rm M_\odot\ \rm yr^{-1}]$ & $1300-5300$ & $560-5900$ & -- \\[0.5mm]
            $\rm\dot{\rm p}_{\rm OF}$ & $[10^{35}\ \rm dyne]$     & $28-110$ &  $19-200$ & -- \\[0.5mm]
            $\rm\dot{\rm E}_{\rm OF}$ & $[10^{43}\ \rm erg \rm s^{-1}]$& $4.7-19$ & $5.0-530$ & --  \\[0.5mm]
            $\rm\tau_{\rm OF}$        & [Myr] & $8.3-33$ & $8.4-90$ & -- \\[0.5mm]
            $\rm\tau_{\rm OF+SFR}$    & [Myr] & $4.9-8.8$ & $6.1-18$ & $23$ \\[0.5mm]
            \enddata
            \tablecomments{Measured properties of the 119 \mum dust continuum (size and flux), OH emission and absorption lines (with formal uncertainties from the fitting procedure), and derived properties of the molecular outflows. }
        \end{deluxetable}
        
\section{Discussion} \label{sec:discussion}
    We begin our discussion by first summarising the known properties of the high--z DSFG sample and their molecular outflows \citep{Spilker2020a,Spilker2020b}. The spatial resolution of their observations were capable of resolving several distinct clumps in the blue--shifted OH absorption, in contrast to the generally smooth background dust continuum. Although not capable of resolving the clumps themselves, \cite{Spilker2020a} use optical depth arguments to predict a substantially more clumpy structure on scales of $\sim50$–200 pc. The comparatively high detection rate of molecular outflows in the sample compared with the measured covering fractions suggest a `fortuitous' geometry of the outflows, such that at least a fraction of the outflow is likely to intervene the line of sight and background continuum, regardless of viewing angle. Moreover, higher outflow covering fractions and higher detection rates are found in sources with higher $\rm L_{IR}$, suggesting that more IR luminous galaxies are capable of driving either more widespread outflows, outflows with larger opening angles, or larger clump sizes. All three scenarios are consistent with the observed increase in FWHM and EW of the outflowing absorption signature towards higher $\rm L_{IR}$ (Fig.\ \ref{fig:ParGrid}). Lastly, they find that the optical depths of the outflowing clumps often peak towards the outskirts of the background continuum, offset from the peak in continuum emission. Whether this is the result of a clumpy expanding shell or the removal of more easily ejected gas at lower column densities found at larger radii, remains unclear. 
    
    \subsection{What Drives Molecular Outflows in Unobscured QSO Hosts?}
        When treated under the same assumptions, the OH molecular outflows observed in J2310+1855 and P183+05, are less or comparably energetic to those driven by the high--z DSFGs (Fig.\ \ref{fig:FIRvsE}). This is despite harbouring comparable SFRs and an additional luminous AGN.
        
        Returning to Fig.\ \ref{fig:ParGrid}, we highlight the key difference in basic host galaxy and blue shifted absorption properties between the QSOs and the DSFG sample. Firstly, both QSOs display narrower than expected absorption signatures for their $\rm L_{IR}$ when compared to the DSFG sample. This may be due to a more cohesive outflow velocity structure (i.e., sheet), fewer velocity components contributing to the overall absorption signature (fewer clumps) or an overall less turbulent outflow. P183+05 consistently deviates towards narrower FWHMs from all of the DSFG relations, but does not significantly deviate in any other way with regards to both outflow and host galaxy properties. J2310+1855, on the other hand, deviates in three additional ways, 1) towards much higher outflow absorption strengths (EW), 2) towards slower outflow velocities and 3) towards a smaller dust continuum size, giving J2310+1855 the highest FIR surface brightness of both samples. We investigate the possible effects of the dust continuum size in Sec. \ref{subsec:bgcont}.
        
        Our upper limits on the outflow $\dot M_{OF}$ and energetics, assuming a covering fraction of 100\% around J2310+1855 and P183+05, provide values within the range where the star formation in these systems is sufficient. Not only is additional energy or momentum from an AGN simply not required to drive these outflows, but the fact that J2310+1855 and P183+05's do not even appear to be boosted with respect to the DSFGs in energetics, $\dot{\rm M}_{OF}$ or velocity, suggests that these outflows may indeed be driven by the same mechanisms -- those associated with star formation. 
        
    \subsection{Interaction Between Central AGNs and the Surrounding ISM}
        At low--z, it is consistently found that higher AGN fractions and luminosities drive faster and more energetic molecular outflows \citep{Sturm2011,Spoon2013,GonzalezAlfonso2017}. Outflow properties also correlate best with AGN luminosity or AGN fraction, whilst only weakly correlating with the star--forming properties of the host galaxy \citep{Calderon2016}. Studies measuring 9.7\mum silicate absorption, which indicates the degree of obscuration originating in the nuclear torus and/or host galaxy disk find strong correlations with the strength of the OH absorption (equivalent width) \citep{Stone2016,Veilleux2013,Spoon2013}. Together, these results suggest that not only the strength of the radiation field, be it from star--formation or an AGN, but the obscuration and thus ability of these energy and momentum sources to couple with the surrounding ISM, is crucial in driving a strong molecular outflow. Indeed \cite{GonzalezAlfonso2017} find in their sample of 14 local Ultraluminous Infrared Galaxies, that the highest molecular outflow velocities traced by OH absorption are found in buried sources, where slower but more massive expansion of the nuclear gas is found. 
        
        But how comparable are low--z AGN hosts compared with our high--z unobscured QSOs? The entire sample studied by \cite{GonzalezAlfonso2017} were galaxies undergoing gas--rich mergers, and indeed AGNs found in the low--z universe are ubiquitously inbedded in such host galaxies \citep{Sanders1996}, where AGN feedback is thought to be stochastically injected into the chaotic surrounding ISM throughout the merging process \citep{GonzalezAlfonso2017}. The three $z>6$ QSOs studied in this paper, however, all display signatures of orderly, rotationally dominated gas reservoirs (Tab. \ref{tab:Litvals}, \citealt{Neeleman2021,Tripodi2022}), opposing a merger scenario. As unobscured QSOs, they have already removed the obscuring material surrounding the AGN, clearing a path through the ISM along the line of sight (note: this need not be aligned with the host galaxy).  
        
        At high--z, the interaction between central AGNs and their host galaxies is a matter under investigation. \cite{RojasRuiz2021} found that strong synchrotron emission extending into mm-wavelengths in the radio-loud QSO PSO J352.4034–15.3373 contributed significantly to its FIR luminosity. This source, however is one of the brightest radio-loud sources in the early universe and the subsequent study of \cite{Khusanova2022}, considering five radio-loud $z>6$ QSOs, found that the SFR distribution and [CII] luminosities of their sources were comparable with the radio-quiet population, suggesting that radio jets to not contribute significantly to these properties.
        
        Significant AGN contributions to the FIR luminosity via dust heating has been measured in some high--z QSOs. \cite{Schneider2015} found an AGN contribution between 30-70\% in the $z=6.4$ QSO J1148+5251. Although chosen as a prototype for the general population of high--z QSO, this source harbours an X-ray luminosity twice that of J2310+1855 and has been detected in CO(17-16) \citep{Gallerani2014} which remains undetected in J2310+1855. \cite{Carniani2019} inspected the CO SLEDs and FIR properties of these sources and concluded that the differences found were likely due to different gas heating mechanisms present, those being mechanisms associated with the central AGN for J1148+5251 and star formation in J2310+1855. \cite{Decarli2022} similarly find, using multi-line diagnostics and the non-detection of high-J CO transitions, that star formation also appears to drive the molecular gas excitation in P183+05. \cite{Duras2017} found an average AGN contribution of 50\% in their sample of 16 IR--bright QSOs. This sample was chosen to represent an intermediate population of QSOs emerging from the heavily obscured phase, but preceding the blue QSO phase where intervening material has already been swept away. This again differs from the population of unobscured, UV and FIR bright QSOs represented by our QSOs.
        
        In a luminosity--limited sample of $z>6$ unobscured QSOs (including our sources), no correlation is seen between quasar UV luminosity ($M_{1450}$, see Tab. \ref{tab:Litvals}) and FIR luminosity \citep{Venemans2018}. This is true even in the central regions surrounding the QSOs \citep{Venemans2020}, suggesting little interaction between the QSOs and host galaxy ISM. Our assumption that the SFR dominates the ${\rm L_{FIR}}$ of our sources is thus likely a good assumption and it is perhaps also unsurprising that we do not find evidence of a boost in the molecular outflow properties measured and derived for our unobscured QSO hosts, compared to the DSFG sample. 
        
        In any case, we can not rule out a contribution from the AGN to the FIR luminosity with this information alone, nor can we ignore the large uncertainties associated with estimating this value \citep{Venemans2018,Venemans2020,Tripodi2022}. Therefore, as an extreme case, we consider an AGN contribution of 50\% to the ${\rm L_{FIR}}$, as found for high--z BAL QSOs \citep{Duras2017}. We expect, for the reasons stated above, that this value is an upper limit of the AGN contribution for our unobscured sources. This effect would move our sources left, by half, in Fig.s \ref{fig:FIRvsMOFR}, \ref{fig:FIRvsE} and in the left two columns of \ref{fig:ParGrid}, if one is to consider only the input from star formation. In Fig. \ref{fig:ParGrid} (note log scales), this makes little difference to our conclusions. In Fig.s \ref{fig:FIRvsMOFR} and \ref{fig:FIRvsE}, P183+05 indeed begins to appear more energetic than expected for its SFR, if it is assumed that all the sources have an outflow covering fraction of 100\%. This moves the upper limit of the energy flux within the `high coupling efficiency or AGN required' region (indicated by the smaller faint orange symbols). We remind the reader that a covering fraction of 100\% is an extreme scenario, with the true values expected to be close to the fractional absorption depth, which in the case of P183+05 is only 9.5\%. Assuming an AGN contribution to the FIR of 50\%, a high coupling efficiency or an AGN contribution to drive the molecular outflow seen in P183+05 is not required unless the outflow covering fraction is $\geq53\%$. This is higher than all of the absorption depths measured by \cite{Spilker2020a} for outflowing or systemic gas in their sample. Thus, whilst it is possible the molecular outflow in P183+05 lies in the parameter space requiring either a high coupling fraction or an AGN contribution, this would require either a significant AGN contribution to the FIR luminosity and/or a high outflow covering fraction ($\geq53\%$). For reasons stated above we believe both the condition of a significant $\rm L_{FIR}$ AGN contribution and/or high covering fraction to move P183 into this parameter space to be unlikely, but do not rule it out. For our other source, J2310+1855, a $\rm L_{FIR}$ 50\% AGN contribution does not display higher than expected values and does not require a high coupling efficiency or AGN contribution even with a 100\% outflow covering fraction. In summary, an AGN contribution to the driving of the molecular outflow in J2310+1855 is not required and is only possibly required in P183 under particular extreme scenarios.
        
        Recent results studying the ionised gas phase, however, have found extremely fast outflows up to 17\% the speed of light in $z\sim6$ QSOs using CIV absorption \citep{Bischetti2022}. Such outflows must transport enormous amounts of energy into the circum-- and intergalactic--medium from the central engine. Included in this sample is J2310+1855, with a detected ionised gas outflow travelling between ~18500-27000 \kms. Although clearly not associated with the molecular outflowing gas travelling at just $\sim300 \kms$, this may suggest a scenario where a combination of star formation driven molecular outflows and AGN driven ionised outflows work in conjunction to quench these massive active galaxies and transfer energy to atomic outflows found at higher radii in the surrounding halos of diffuse turbulent gas (as found in J2310+1855 in \OHp \citep{Shao2022}). As seen in low--z studies, molecular gas outflows dominate the mass and momentum transport of matter out of the galaxy, whilst the warmer ionised phases dominate the kinetic energy budget \citep{Fluetsch2021}. The molecular outflows presented here are indeed capable of transporting large amounts of gas, providing short depletion times on the order of Myrs predicted for these galaxies \citep{Simpson2012,Costa2018}. As discussed in Sec. \ref{sec:escfrac}, however, a negligible fraction of the outflowing molecular gas will escape the gravitational bind of their host galaxy, and will be re--accreted unless additional energetic input is provided at higher radii. Thus, whilst the molecular phase may predominantly remove the fuel for future star formation directly from the galaxy, the ionised phase may be responsible for keeping that material at higher radii by efficiently injecting energy into the CGM and thus providing a scenario where these massive gas rich galaxies can quench on short timescales.
        
        During the QSOs' embedded or blowout phase, however, trapped radiation from the central engine may play a much greater role in launching molecular outflows \citep{Costa2018}. Observing QSOs in these early stages will therefore be crucial in our studies determining the full impact of QSOs on their surrounding ISMs and the role they play in ejecting gas from the galaxy.
        
    \subsection{The Background Dust Continuum and its Effect on the Outflow Absorption Signature}\label{subsec:bgcont}
        The conclusion of the previous subsection does not, however, explain the differences in spectral properties of the OH outflows observed between our $z>6$ QSO and the high--z DSFGs. The combination of narrower line widths and comparable or higher EWs of the total outflow absorption may imply an increase in outflow covering fraction or clump positions that preferentially block brighter regions of the background dust emission. Both scenarios may be true in the case of J2310+1855 whose small 119 \mum dust size and high surface brightness (Fig.\ \ref{fig:ParGrid}) means any clumps intervening the line of sight will have the tendency to cover brighter regions, and will block a larger fraction of the overall surface area. For galaxy disks at higher inclinations in the observer's line of sight, such effects of intervening clouds are further intensified.
        
        \begin{figure}
            \plotone{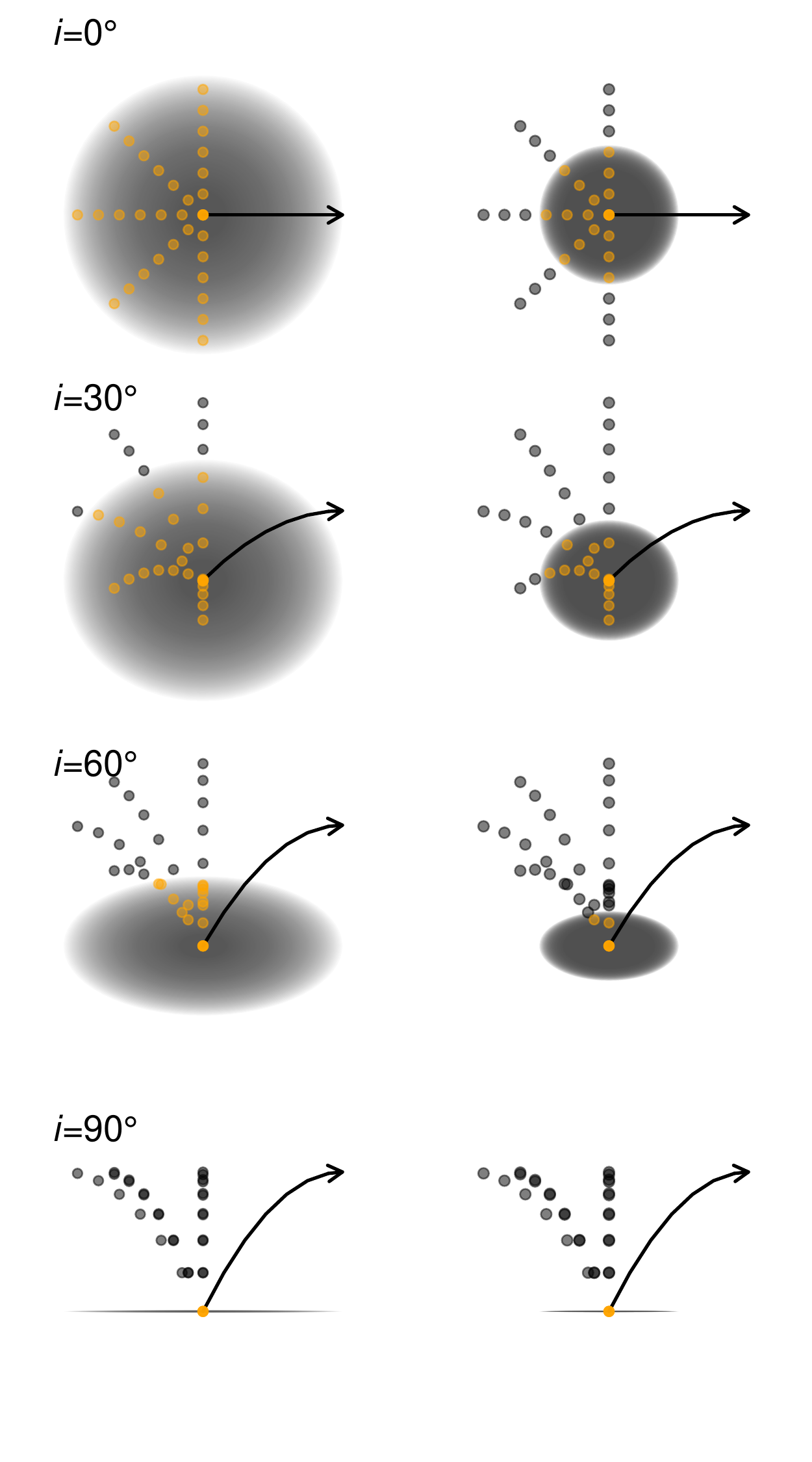}
            \caption{Idealised schematic of two equally luminous disky galaxies (left and right) seen at four different inclinations (from the top: $0^\circ$, $30^\circ$, $60^\circ$ and $90^\circ$). We consider a clumpy outflow ejected from the center of the galaxy, travelling radially in five angles covering the left hand side of the disk, and vertically through a galactic fountain trajectory. On the right we show this trajectory as a solid black arrow. Orange circles indicate clumps that intervene the observer's line of site and the background continuum, and thus contribute to the absorption signature of the outflow. Grey circles indicate when clumps do not intervene and are invisible to the observer via absorption.  \label{fig:cartoon}}
        \end{figure} 
        
        We illustrate this effect in Fig.\ \ref{fig:cartoon}, for two galaxies of equal luminosity but different size (different surface brightness), observed at four different inclinations. We consider the effect of the background continuum size and inclination on the observability of foreground clumps travelling in a galactic fountain trajectory, ejected from the center of the galaxy. We show this trajectory for five radial directions but note that clumps may be ejected from anywhere in the galactic plane and travel outwards in any direction. Clumps that intervene the observer's line of sight and background dust continuum and are therefore observable via absorption are coloured orange. We find that for smaller disk sizes and higher inclinations, absorption line observations become limited to clumps at low radial distances or those lying more towards the foreground. Individual clumps in these circumstances cover a greater fraction of the background dust area, tend to block regions of higher surface brightness and thus imprint a stronger absorption signature on the global spectrum. Fewer clumps are therefore needed to absorb the same or more background emission, limiting the number of velocity components contributing to the velocity dispersion of the global absorption line.
        
        Without resolving the molecular outflow we can not say for certain what the geometry or trajectory of the outflowing gas is. Fig.\ \ref{fig:cartoon} does, however, illustrate that clumps at smaller scale heights are preferentially observed. If a clump is initially ejected perpendicular from the disk (the center or farther out), and then continues along a trajectory that deviates from this initial direction, (such as the fountain scenario depicted in Fig.\ \ref{fig:cartoon}), the restriction of our observations to low altitude clumps also restricts our observations to clumps with higher perpendicular velocities. The combined outflow velocity therefore becomes more dependent on the inclination of the disk. 
        
        Thus, the smaller disk size of J2310+1855 may therefore cause the absorption signature to be more sensitive to projection effects, contributing to the offset in outflow velocity shown in the first panel of Fig.\ref{fig:ParGrid}. Neither J2310+1855 or P183+05, however, are found to have significantly inclined disks ($25$ and $<22$ deg), and so it is not expected that projection effects play a dominating role in the slower than expected outflow velocities. We note, however, that \cite{Tripodi2022} found similar results for their dynamical models of J2310+1855 for inclinations over the range of [20, 45] deg and earlier dynamical modelling by \cite{Feruglio2018} using lower resolution data have previously found a higher inclination of 53 deg. \cite{Tripodi2022} discarded models with $i>30$ deg due to the gas mass exceeding the dynamical mass, however, this cut off is dependent on the reliability and suitability of the CO conversion factor for a high--z unobscured QSO. In the case of P183+05, which does not display an unusually small continuum size and has only an upper limit on the host galaxy inclination, it is unclear what causes the smaller than expected FWHM of the absorption line and we do not rule out simple natural variance.
        
        Small QSO host dust sizes have been measured in a number of samples at high--z \citep{Ikarashi2015,DAmato2020,Venemans2020,Stacey2021}, although evidence for an additional shallower extended component is found in the stacking of 27 $z\geq6$ QSO hosts \citep{Novak2020}. \cite{Stacey2021} divided their z=1.5-2.8 QSO hosts into two categories: those exhibiting a clumpy dust distributions, with sizes and star formation rate densities typical of sub--mm--selected DSFGs, and those  with no evidence of clumpy features and characterised by compact ($\rm R_{eff}<1\ kpc$) sizes and high star formation densities. The latter category is believed to indicate a phase of dissipative contraction before becoming compact spheroids, and describes both the QSO hosts presented in this paper and the DSFGs presented by \cite{Spilker2020a}. The optical/infrared sizes of QSO hosts at $z\sim1.5$ \citep{Silverman2019}, place them between main sequence and quiescent galaxies, whilst \cite{Ikarashi2015} found in their sample of $z\sim1$–3 QSOs and starbursts, that composite sources harbour the smallest sub--mm dust sizes. While all three of our QSO sources lie within the range of dust continuum sizes of the DSFG comparison sample, they do lie at the extreme end of the surface brightness. In particular, J2310+1855 exhibits a much smaller continuum size for its infrared luminosity and the highest surface brightness of the two samples, possibly indicating that this source is in a more advanced stage in the transformation into a quiescent spheroidal, consistent with the shorter depletion timescale estimated for this source. We note again that our size estimate of P036+03 was significantly larger than previous measurements \citep{Venemans2020} which would instead make P036+03 our most compact source, and possibly also further along in its dissipative collapse. This may also provide an explanation for the absence of OH 119 \mum absorption in P036+03, as outflows from a more evolved system may have travelled farther away from the bright central region, and a more compact background emission decreases the total area in which an outflow may be observed. The combination of these effects would decrease the likelihood of an outflowing component intercepting the line of sight with the background continuum.
        
    \subsection{OH 119 \mum Emission}\label{subsec:OHemission}
        OH 119 \mum emission has been detected in many of the low--z galaxy samples \citep{Sturm2011,Veilleux2013,Spoon2013,Stone2016,Calderon2016}, and is typically associated with the presence of an AGN. AGN fraction, and not simply AGN luminosity, was found to be key in setting the OH 119 \mum character (relative strength of emission to absorption) \citep{Veilleux2013,Stone2016}. In the most AGN dominated systems where OH is seen purely in emission, \cite{Veilleux2013} notes that the line--widths remain modest, possibly indicating a phase after which the galaxy has cleared away the obscuring molecular gas. 
        
        Deeper 9.7 \mum silicate absorption (a measure of obscuration in both nuclear torii and host galaxy disks) has been found to correlate with fainter OH 119 \mum emission and deeper OH 119 \mum absorption, indicating that OH 119 \mum emission is strongly affected by the geometry of the obscuring material \citep{Stone2016}. Whilst earlier studies argued that OH 119 \mum emission originates from the buried nuclear regions of galaxies (e.g., \citealt{Spoon2013}), where radiative pumping would predominantly excite the OH, \cite{Stone2016} later showed that in some Type 2 AGNs (where the nucleus is obscured), and where the silicate 9.7 \mum line is found in absorption, some of the OH 119 \mum is still found in emission. Given the apparent impact of obscuring material on OH 119 \mum emission this may imply that the OH 119 \mum emission also originates farther out than the nucleus, such as in the circumnuclear starburst. Here, physical conditions favor collisional excitation over radiative pumping. Similar conclusions have emerged from a studies measuring the relative strengths of OH transitions (e.g. in NGC 1068 \citealt{Spinoglio2005}) and in studies using both silicate absorption and multi OH transition data \citep{Runco2020}. 
        
        OH 119 \mum emission was not detected in any of the \cite{Spilker2020a} high--z DSFGs, consistent with their measured subdominant AGN fractions (all upper limits). OH 119 \mum emission is likely present in the z = 6.13 QSO ULAS J131911+095051 \citep{HerreraCamus2020}, causing the blue--shifted absorption observed in it's spectra to exhibit only a single absorption peak, due to superimposed emission and absorption of the low and high frequency doublet transitions, respectively. A similar scenario is more clearly observed in P183+05 of our sample. Tentative emission is also detected in a second of our sources, P036+03, but not in that of J2310+1855, which instead displays the strongest OH absorption. Thus, as seen at low--z, it appears the presence of an AGN in high--z sources is also required to provide an environment in which OH 119 \mum emission may be observed. Also consistent with low--z observations, emission line--widths appear to be modest.
        
        All three of our QSOs are unobscured and as mentioned in our previous discussion sections, it is suspected that the outflowing OH observed in absorption is ejected via star formation feedback and thus likely not to have originated from the nucleus region. It may then be expected that the gas traced by OH 119 \mum emission, is also, at least in part, located farther out from the AGN, perhaps in a circumnuclear starburst as predicted for some low--z galaxies \citep{Spinoglio2005,Spoon2013,Runco2020}. Furthermore, we find in the case of P183+05, that the OH 119 \mum emission is spatially more extended than the background continuum. We can not however determine with a single OH transition what excitation mechanism is dominant in these sources. Future multi--line and spatially resolved OH observations will be needed to elucidate the nature and location of these environments. 
        
\section{Conclusion} \label{sec:conclusion}
     We present marginally resolved ALMA band 7 observations targeting the OH 119 \mum line and dust continuum in three $z>6$ QSOs: J2310+1855, P183+05 and P036+03. We detect OH absorption in two of the sources (J2310+1855 and P183+05), blue--shifted with respect to the host galaxy systemic velocities, and in emission in P183+05 and tentatively P036+03, at systemic velocities. We detail our main results below.
     
     \begin{itemize}
         \item We compare the spectral properties of the blue--shifted OH 119 \mum absorption signatures tracing the molecular outflows in J2310+1855 and P183+05, with those of a comparison high--z DSFG sample \citep{Spilker2020a,Spilker2020b}. We find that the FWHM in both J2310+1855 and P183+05 are narrower than expected for their host galaxy FIR luminosities, but that the equivalent width (absorption strength) of the line is significantly larger in the case of J2310+1855, and comparable in P183+05. This may imply a higher outflow covering fraction or a geometry of the outflow that has a tendency to cover brighter regions of the background dust continuum.
         
         \item Assuming a spherical thin shell geometry, we derive masses, mass outflow rates and energetics of the molecular outflows. We find that J2310+1855 and P183+05 have comparable mass outflow rates to the DSFG sample for the $\rm L_{FIR}$. J2310+1855 is significantly offset towards lower energy flux for it's $\rm L_{FIR}$.
         
         \item Whilst the estimated escape fractions of the molecular outflows are negligible, we suggest that energy injected into the CGM via the fast nuclear--driven ionised phase may \citep{Bischetti2022} may inhibit the re-accretion of this gas back onto the galaxy. This brings our derived depletion times of $<1$ Myr into agreement with predictions for massive high--z QSOs \citep{Simpson2012,Costa2018}.
         
         \item We do not find any evidence, nor require any additional input from the central AGN to drive the observed molecular outflow in J2310+1855 but can not rule out an AGN contribution in P183+05 if the AGN contribution to $\rm L_{FIR}$ is significant and/or the outflow covering fraction is high ($\geq53\%$), which we find to be an unlikely scenario for our $z>6$ unobscured QSOs \citep{Venemans2018,Venemans2020}. This suggests that the molecular outflows are driven dominantly by processes associated with star--formation, as is expected of the high--z DSFG sample.
         
         \item Consequently, we suggest that the observed spectral differences of the absorption lines are instead caused by differences in the background dust continuum. We demonstrate with an outflow toy model (Fig.\ \ref{fig:cartoon}) that for a more compact and/or inclined background dust continuum, observations are limited to fewer outflowing clumps and clumps at low altitudes. This results in narrower absorption line widths (fewer velocity components) and an increase of the projection effects on the outflow velocity, assuming clumps move in a galactic fountain--like trajectory. For both an increase in inclination, and more compact dust distribution, each individual intervening clump will cover a larger fraction of the overall dust continuum and will have the tendency to cover regions of brighter dust continuum surface brightness. We suggest that J2310+1855's small dust size may make its outflow velocity more sensitive to projection effects, contributing to its slower than expected outflow velocity.
         
         \item J2310+1855's compact dust continuum may indicate that its host galaxy is at a more evolved stage of dissipative contraction in its transformation into a compact spheroidal \citep{Stacey2021}. The small dust sizes measured in other high--z QSO samples suggest that the effects of the background continuum on the absorption line spectral properties of outflows that we find in this paper, may be a common systematic in future, larger samples.
         
         \item We detect OH 119 \mum emission in P183+05, and tentatively in P036+03, in contrast with the high--z DSFG sample where emission is not detected in any of the sources. This is consistent with detections in low--z studies that have found the presence of an AGN is required to provide the necessary environment to excite the molecule. 
     \end{itemize}
     
     Given the limited sample size of three in this study, and the marginally resolved nature of our observations, we note that many of our conclusions are speculative at this time. To concretely determine the geometry and covering fractions of the outflowing molecular gas, spatially resolved observations are crucial. Larger sample sizes will also be required to determine if the differences in outflow spectral properties driven by QSO hosts, as found in this paper, are truly caused by background dust distribution and not the outflow itself. Similarly, larger samples of DSFGs, particularly at higher ($\rm 10^{13}\ L_{\odot}$) $\rm L_{FIR}$, are needed to constrain the trends found in the purely star-forming sample, so as to better compare between the two possible outflow driving mechanisms.
     
     Lastly, in this paper we have focused on three unobscured QSOs, sources that have already completed their blow--out phase and where energy and momentum injected by the central black hole may efficiently escape the galaxy via the cleared out pathway. Our results suggest that at this evolutionary phase the interaction between QSO and host galaxy ISM is limited. Thus observations targeting QSOs in the lead up to or during the blow--out phase may better illuminate the role and impact QSOs play in removing cool gas from their host galaxies.
     
\begin{acknowledgments}
This paper makes use of the following ALMA data: ADS/JAO.ALMA\#2018.1.01790.S. ALMA is a partnership of ESO (representing its member states), NSF (USA) and NINS (Japan), together with NRC (Canada), MOST and ASIAA (Taiwan), and KASI (Republic of Korea), in cooperation with the Republic of Chile. The Joint ALMA Observatory is operated by ESO, AUI/NRAO and NAOJ. This work benefited from the support of the project Z-GAL ANR-AAPG2019 of the French National Research Agency (ANR). Finally, we thank the anonymous referee for their feedback which helped improve this paper.
\end{acknowledgments}

%

\vspace{5mm}
\facilities{ALMA}


\software{\texttt{CASA} (v4.53; \citealt{McMullin2007})}





\bibliographystyle{aasjournal}
\bibliography{OH119z6QSOsOutflows}{}



\newpage
\thispagestyle{empty} 
\vspace*{-72pt}
\noindent\makebox[\textwidth]{\includegraphics[page=1]{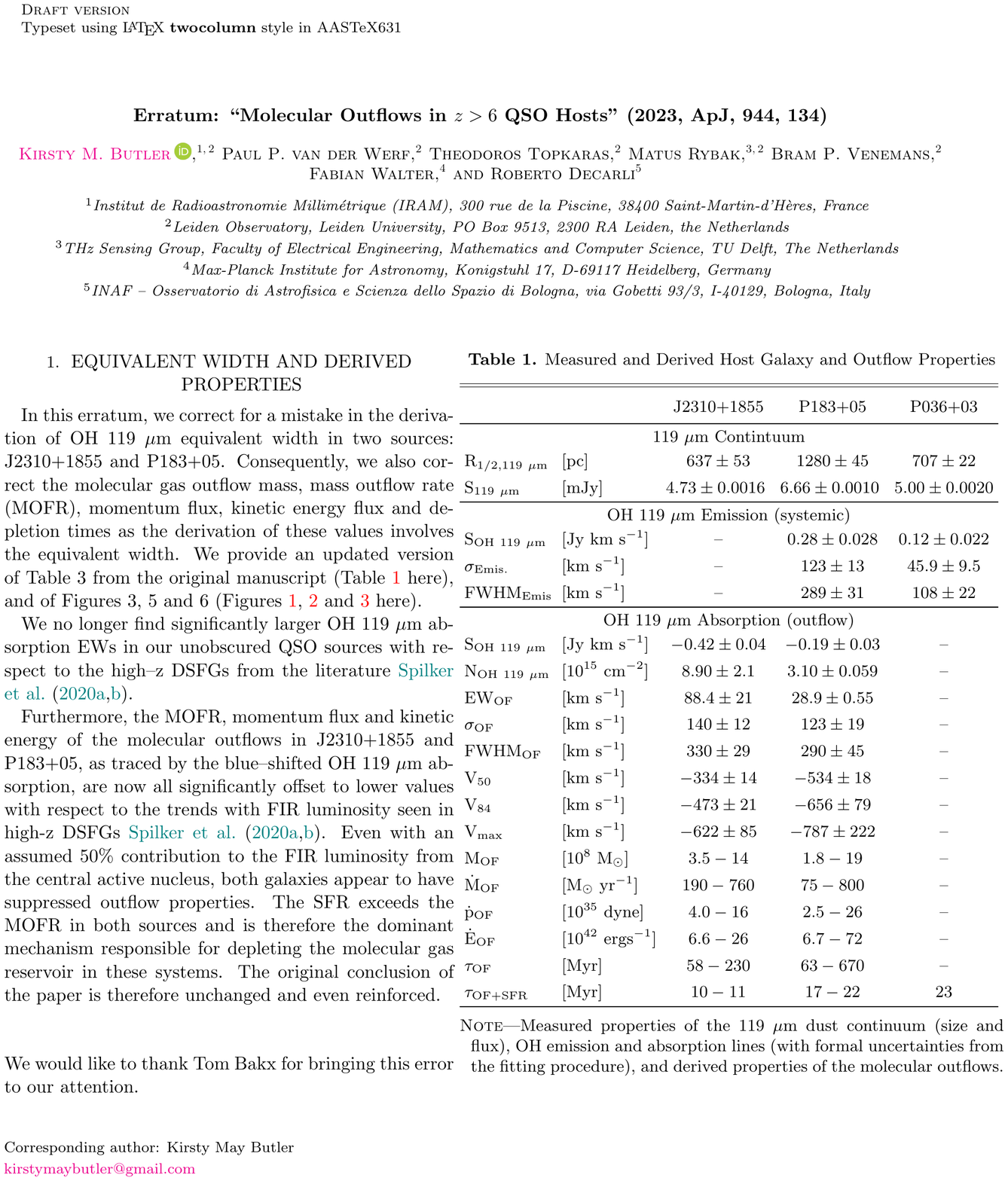}}

\newpage
\thispagestyle{empty} 
\vspace*{-72pt}
\noindent\makebox[\textwidth]{\includegraphics[page=2]{ErratumJPEG}}

\newpage
\thispagestyle{empty} 
\vspace*{-72pt}
\noindent\makebox[\textwidth]{\includegraphics[page=3]{ErratumJPEG}}

\newpage
\thispagestyle{empty} 
\vspace*{-72pt}
\noindent\makebox[\textwidth]{\includegraphics[page=4]{ErratumJPEG}}
\thispagestyle{empty} 
 
\end{document}